\documentclass[aps,prd,preprint,groupedaddress,preprintnumbers,longbibliography,nofootinbib]{revtex4-1}

\newcommand{\vvev}[1]{\left\langle\kern-0.3em\left\langle #1
    \right\rangle\kern-0.3em\right\rangle}

\usepackage{amsmath}
\usepackage{graphicx}
\usepackage{amsfonts}
\usepackage{caption}
\usepackage{subcaption}
\usepackage{soul}
\usepackage{xcolor}
\usepackage{empheq}
\newcommand*\widefbox[1]{\fbox{\hspace{2em}#1\hspace{2em}}}

\begin{document}


\title{A look at the operator product expansion in critical dynamics}


\author{C.~Pagani}
\email[]{cpagani@uni-mainz.de}
\author{J.~Sobieray}\email[]{jasobier@uni-mainz.de}
\affiliation{Institute f\"{u}r Physik (WA THEP)
  Johannes-Gutenberg-Universit\"{a}t\\ Staudingerweg 7, 55099 Mainz,
  Germany}
%


\date{\today}

\begin{abstract}
We consider the critical relaxation of the Ising model, the so-called model A, and study its operator product expansion. Within perturbation theory, we focus on the operator product expansions of the two-point function and the response function. 
At the fixed point, we normalize the coefficients and the scaling variables so that the result displays universality. The role of the fluctuation-dissipation theorem is also discussed, and it is shown that it provides non-perturbative relations among the operator product expansion coefficients. 
Finally, the large $N$ limit is considered.
\end{abstract}


\maketitle

\section{Introduction}

The operator product expansion (OPE)
\cite{Wilson:1969zs,Zimmermann:1972tv}
offers a non-perturbative tool
to study quantum and statistical field theories. The paradigmatic
example is given by two-dimensional critical phenomena, which have
been studied in detail also thanks to the application of conformal
field theory (CFT) techniques in which the OPE plays a key role \cite{Francesco1997a}.
More recently, the conformal bootstrap approach allowed to obtain
precise predictions also in higher dimensions \cite{Poland2019a}. These
results have been obtained by exploiting also the powerful constraints
due to conformal symmetry, which, however, may not always be present
since, for instance, one may not exactly be at criticality or because
of the presence of some external scale, as it happens in certain
non-equilibrium systems. 

However, as such, the OPE holds also outside the critical region,
i.e., away from a fixed point, and it can be employed to study a wide
variety of systems. In particular, the OPE is used to study the expansion
of the two-point function of equilibrium systems close to criticality,
see, e.g., \cite{Brezin:1974zz,Brezin1974b,Caselle:2015csa,Costagliola:2015ier,Caselle:2016mww}. 
In the context of QCD, it is employed to study
deep inelastic scattering processes \cite{Novikov:1977dq}.

Given that the OPE offers such a powerful framework in the description of different physical phenomena,
it is natural to investigate its application in the context of critical dynamics.
The purpose of this paper is to study the operator product expansion
in the context of critical dynamics in the case where the fluctuation-dissipation theorem applies for the first time. 
We shall work within the
response field formalism \cite{Martin:1973zz,Janssen:1976a,deDominicis:1976a},
in which the so-called response
field $\bar{\phi}$ is introduced over and above the physical field
$\phi$. This makes the formalism very similar to what is employed
in genuinely non-equilibrium systems and so it constitutes a first
step towards a general study of the OPE in non-equilibrium systems.
In the context of critical dynamics,
one of the main motivations behind the present work is indeed to perform some basic analysis and some concrete calculations of OPE coefficients,
which has been far less explored than its static counterpart. 
As such, this work has an exploratory field theoretic character.
Let us note that a sort of short-time expansion
is sometimes invoked in the studies concerning the so-called critical
initial slip, which is related to initial states that are not the
equilibrium ones, see \cite{Taeuber_2014} and references therein. 
Furthermore, as time and space scale differently, our results are somehow reminiscent of the OPE for non-relativistic theories, see \cite{Golkar:2014mwa,Shimada:2021xsv}. 
Finally, let us anticipate that,
besides the field theoretic aspects, 
we shall show that
the OPE allows one to define further universal quantities,
besides the critical exponents,
which generalize the OPE coefficients of equilibrium
statistical field theories at criticality.

The paper is organized as follows. In section \ref{sec:Path-integral-approach-model-A}
we introduce our notation and conventions and briefly review some
of the main features of the path integral approach to model A. In
section \ref{sec:Composite-operators} we introduce some composite
operators of interest and analyze their renormalization and expectation
value (VEV). We also briefly consider the consequences of the time-reversal
symmetry at the origin of the fluctuation-dissipation theorem. In
section \ref{sec:Gaussian-OPE} we derive the Gaussian, i.e., free-field
theory, OPE. We do so by considering the functional ``Schroedinger''
picture associated with the Fokker-Planck equation. In section \ref{sec:corrections-to-OPE}
we compute the leading order corrections to the most singular parts
of the OPE of the two-point and the response functions. Finally, in
section \ref{sec:Universal-quantities-from-OPE} we discuss how to
extract universal information from the OPE calculated in the previous
section and derive a universal form of the OPE coefficients to $O\left(\epsilon\right)$.
In sections \ref{sec:Extention-to-ON} and \ref{sec:Large-N-limit}
we consider the $O\left(N\right)$ model both perturbatively and in
the large $N$ limit. We summarize our findings in section \ref{sec:Summary-and-conclusions}.
Some definitions and technical discussions are confined to several Appendices.
We highlight graphically the main results of our work by encircling the most relevant expressions in a box.

\section{Path integral approach to model A \label{sec:Path-integral-approach-model-A}}

In this section, we state our conventions and recall the main features of model A.
For a review, we refer to \cite{Zinn_book}.

\subsection{Model A and its path integral representation}

The dynamics of model A is defined by the following stochastic differential
equation (SDE)
\begin{eqnarray}
\partial_{t}\phi_{0}\left(t,x\right) & = & -D_{0}\frac{\delta H}{\delta\phi_{0}\left(t,x\right)}+\zeta\left(t,x\right)\,,\label{eq:SDE-model_A}
\end{eqnarray}
with $\zeta$ being a Gaussian noise with zero average and covariance
given by
\begin{eqnarray*}
\langle\zeta\left(t,x\right)\zeta\left(t^{\prime},x^{\prime}\right)\rangle & = & 2D_{0}\delta\left(t-t^{\prime}\right)\delta\left(x-x^{\prime}\right)\,.
\end{eqnarray*}
We consider the SDE (\ref{eq:SDE-model_A}) in the Ito sense. The
Hamiltonian $H$ is defined to be that of the Landau-Ginzburg model:
\begin{eqnarray}
H & = & \int_{x}\left[\frac{1}{2}\partial_{i}\phi_{0}\partial_{i}\phi_{0}+\frac{1}{2}m_{0}^{2}\phi_{0}^{2}+\frac{1}{4!}g_{0}\phi_{0}^{4}\right]\,,\label{eq:equilibrium_H_model_A}
\end{eqnarray}
where $\partial_{i}\equiv\partial/\partial x^{i}$. The zero in the
subscript of both couplings and fields signals that those are bare
(non-renormalized) quantities. Our conventions and notations for the
integrals and the Fourier transform are given in Appendix \ref{sec:Conventions}.

Associated with the stochastic process that describes the dynamics
of model A, i.e., the SDE (\ref{eq:SDE-model_A}), there is the following
Fokker-Planck equation \cite{Zinn_book}:
\begin{eqnarray}
\partial_{t}P\left[\phi\right]\left(t\right) & = & D\Biggr\{\int_{x}\frac{\delta}{\delta\phi\left(x\right)}\left(\frac{\delta H}{\delta\phi\left(x\right)}\right)+\int_{x}\frac{\delta^{2}}{\delta\phi\left(x\right)^{2}}\Biggr\} P\left[\phi\right]\left(t\right)\nonumber \\
 & \equiv & -H_{{\rm FP}}P\left[\phi\right]\left(t\right)\,,\label{eq:functional_FP_eq_and_FP_Hamiltonian}
\end{eqnarray}
which implicitly defines the ``Fokker-Planck Hamiltonian'' $H_{{\rm FP}}$.

The stochastic process (\ref{eq:SDE-model_A}) has also an associated
path integral representation, which has been introduced thanks to
the works of Martin, Siggia, Rose, Janssen, and De Dominicis \cite{Martin:1973zz,Janssen:1976a,deDominicis:1976a}.
Such a path integral is constructed by introducing a further field,
which is referred to as the ``response field'' and is denoted by $\bar{\phi}$.
Thus, one introduces the following generating functional (see, e.g.,
\cite{Taeuber_2014}):
\begin{eqnarray*}
Z\left[J_{\phi},J_{\bar{\phi}}\right] & \equiv & \int{\cal D}\phi{\cal D}\bar{\phi}\,\exp\left[-S\left[\phi,\bar{\phi}\right]+\int_{t,x}J_{\phi}\left(t,x\right)\phi\left(t,x\right)+\int_{t,x}J_{\bar{\phi}}\left(t,x\right)\bar{\phi}\left(t,x\right)\right]\,,
\end{eqnarray*}
where the action is given by
\begin{eqnarray}
S & = & \int_{t,x}
\left[
\bar{\phi}_{0}\left(t,x\right)\left(\partial_{t}\phi_{0}\left(t,x\right)+D_{0}\frac{\delta H}{\delta\phi_{0}\left(t,x\right)}\right)-D_{0}\bar{\phi}_{0}\left(t,x\right)^{2}\right]\,.\label{eq:action_model_A-1}
\end{eqnarray}

It is important to note that the above action is invariant under the
following time reversal symmetry (TRS) transformation:
\begin{eqnarray}
 & \begin{cases}
\phi_{0}\left(t,x\right)\rightarrow\phi_{0}\left(-t,x\right)\\
\bar{\phi}_{0}\left(t,x\right)\rightarrow\bar{\phi}_{0}\left(-t,x\right)-\frac{1}{D_{0}}\dot{\phi}_{0}\left(-t,x\right)
\end{cases},\label{eq:TRS-bare-fields}
\end{eqnarray}
where the dot indicates a time derivative. This symmetry implies the
so-called fluctuation-dissipation theorem (FDT), which relates the
two-point function with the response function. 
For a more detailed
discussion of these aspects, we refer the reader to 
\cite{Taeuber_2014,Canet:2006xu,Sieberer:2015hba}.

\subsection{Perturbative renormalization of model A in $4-\epsilon$ dimensions}

By inserting the Hamiltonian (\ref{eq:equilibrium_H_model_A}) in
(\ref{eq:action_model_A-1}) one obtains the following action
\begin{eqnarray}
S & = & \int_{t,x} 
\Biggr\{
\bar{\phi}_{0}\left(t,x\right)
\Biggr[
\partial_{t}\phi_{0}\left(t,x\right)+D_{0}\left(-\partial^{2}+m_{0}^{2}+\frac{g_{0}}{3!}\phi_{0}\left(t,x\right)^{2}\right)\phi_{0}\left(t,x\right)
\Biggr]
\nonumber \\
 &  & -D_{0}\bar{\phi}_{0}\left(t,x\right)^{2}\Biggr\}\,.\label{eq:action_model_A-2}
\end{eqnarray}
By expressing (\ref{eq:action_model_A-2}) in terms of the renormalized
parameters one writes
\begin{eqnarray}
S & = & \int_{t,x}
\Biggr\{
Z_{\bar{\phi}}^{1/2}Z_{\phi}^{1/2}\bar{\phi}\left(t,x\right)
\Biggr[
\partial_{t}\phi\left(t,x\right)+Z_{D}D\left(-\partial^{2}+Z_{m^{2}}m^{2}+\frac{Z_{g}g}{3!}Z_{\phi}^{1/2}\phi\left(t,x\right)^{2}\right)\phi\left(t,x\right)
\Biggr] 
\nonumber \\
 &  & -Z_{\bar{\phi}}Z_{D}D\bar{\phi}\left(t,x\right)^{2}\Biggr\}\,.\label{eq:action_model_A-3}
\end{eqnarray}

The model is perturbatively renormalizable \cite{Taeuber_2014} and we shall
work with the minimal subtraction scheme in dimensional regularization
so that the counterterms $Z_{i}-1$ can be expressed as a sum of poles
in $\epsilon$. It should be noted that the renormalization factors
$Z_{i}$ are not fully independent from one another, they are related
by the fluctuation-dissipation theorem, i.e., the symmetry (\ref{eq:TRS-bare-fields}).
The renormalization factors appearing in the action (\ref{eq:action_model_A-3})
have been calculated many times and we refer the reader to \cite{Taeuber_2014}
for a review. 

Moreover, since in the static limit one recovers the equilibrium distribution,
the renormalization, 
and so the assiociated critical exponents, 
of the couplings appearing in the Hamiltonian (\ref{eq:equilibrium_H_model_A}) are
the same as in the static case. However, when discussing dynamical
critical systems, a further exponent arises, the so-called dynamical
critical exponent $z$, which describes the relation between spatial
and temporal scalings and is related to the renormalization of the
coupling $D$: $z=2+\gamma_{D}$ with $\gamma_{D}=D^{-1}\mu\frac{d}{d\mu}{D}=-Z_{D}^{-1}\mu\frac{d}{d\mu}{Z}_{D}$.
At the Gaussian fixed point one has $z=2$, which implies that a two-point
function takes the following scaling form: $G\left(t,x\right)=x^{-2\Delta}g\left(\frac{t}{x^{2}}\right)$.

Finally, let us assign a (bare)
scaling dimension to each field. The scaling dimension of the field
$\phi$ is $\frac{d-2}{2}$, i.e., $\phi$ scales like $x^{-\frac{d-2}{2}}$
while the scaling dimension of the response field $\bar{\phi}$ is
$\frac{d+2}{2}$, i.e., $\bar{\phi}$ scales like $x^{-\frac{d+2}{2}}$.

\newpage
\section{Composite operators \label{sec:Composite-operators}}

The operator product expansion allows one to express the insertion of two composite operators in a correlation function in the short distance limit
as a sum of insertions of a single composite operator multiplied by suitable coefficients, the so-called Wilson coefficients \cite{Wilson:1969zs,Zimmermann:1972tv}. 
In practice, the OPE states the validity of
\begin{equation}
\left[O_i \left(x\right) \right]
\left[O_j \left(y\right) \right] =
\sum_k C_{ijk}\left(x-y\right) \left[O_k \left(\frac{x+y}{2}\right) \right]\, ,
\label{eq:def_OPE_statement}
\end{equation}
inserted into any correlation function in the limit of $\left|x-y \right|$ approaching zero.
In (\ref{eq:def_OPE_statement}), $C_{ijk}\left(x-y\right)$ indicates a Wilson coefficient and the operators $\left[O_i \right] \left[O_j \right]$
are taken to be time ordered.

It must be emphasized that the composite operators $\left[O_i\right]$ are defined as renormalized composite operators, i.e., 
they are not mere field monomials, 
such as $\phi^2$,
but they are modified by suitable renormalization factors.
In this section, within perturbation theory,
we consider the first order renormalization of certain
composite operators, which we shall use later on to compute the most singular parts of the
OPE of the two-point function and the response function.
In the case of the two-point function, 
one has 
$\left[O_i\right]=\left[O_j\right]=\phi$ 
while in the case of the response function 
$\left[O_i\right]=\phi$
and $\left[O_j\right]=\bar{\phi}$. 
We also determine certain normalizations factors of interests.

Let us start by making few general considerations. We shall work in
$d=4-\epsilon$, within perturbation theory, and adopt dimensional
regularization together with minimal subtraction to renormalize composite
operators. In this setting, in order to define a composite operator
$\left[O\right]$, one has that the ``naive'' operator $O$ requires
mixing with operators of the same (or smaller) scaling dimension,
schematically $\left[O_{i}\right]=Z_{ij}O_{j}$. For example, the
operator $\phi^{2}$ can mix only with the identity, $m^{2}\mathbb{I}$,
whereas the operator $\phi^{4}$ mixes with $\phi\bar{\phi}$, among
other operators. 
This latter example tells us that, 
in the dynamical extension of the Landau-Ginzburg Hamiltonian, 
the composite operators require new mixing with respect to the static case, which clearly
does not possess any mixing with operators depending on the response field.\footnote{
Let us note that in principle one could integrate out the response
field and work with an Onsager-Machlup kind of action 
\cite{OnsagerMachlup1953}.
In this setting, one would not have any mixing with $\bar{\phi}$-dependent
operators since the response field has been integrated out. However,
throughout this work, we shall always keep the response field since
we are interested also in the response function.}

\subsection{The composite operator $\left[\phi^{2}/2\right]$ \label{subsec:The-composite-operator-phi2}}

We wish to construct the renormalized composite operator $\left[\phi^{2}/2\right]$
to order $O\left(\epsilon\right)$. The composite operator takes the
form:
\begin{eqnarray*}
\left[\frac{\phi^{2}}{2}\right] & \equiv & Z_{22}Z_{\phi}\frac{\phi^{2}}{2}+Z_{20}m^{2}\,,
\end{eqnarray*}
where actually the wave function renormalization is negligible at
this order. 
We shall limit ourselves to list some results and refer the redear to Appendix \ref{app:details_phi2} for more details.

We begin our discussion by considering the VEV of
$\left[\phi^{2}/2\right]$
and its expression at tree level.
Let us write it down explicitly within
the context of dimensional regularization:
\begin{eqnarray}
\Bigr\langle\left[\frac{\phi^{2}}{2}\right]\Bigr\rangle\,=\,\Bigr\langle\left(\frac{\phi^{2}}{2}+\frac{m^{2}\mu^{-\epsilon}}{16\pi^{2}\epsilon}\right)\Bigr\rangle+O\left(g\right) & = & \frac{m^{2}}{32\pi^{2}}\left(\gamma-1+\log\frac{m^{2}}{4\pi\mu^{2}}\right)+O\left(\epsilon\right)\,,
\end{eqnarray}
where we do not display explicitly the $O\left(\epsilon\right)$-terms,
which however should be kept.
As far as the scaling dimension of the operator is concerned,
one obtains the following expression
to the first order in the $\epsilon$-expansion:
\begin{eqnarray}
\Delta_{\phi^{2}} & = & d-2+\gamma_{\phi^{2}}\,=\,2-\frac{2}{3}\epsilon\,.
\label{eq:Delta_2_O_epsilon}
\end{eqnarray}

Finally, let us also consider the static limit of the two-point of
$\left[\phi^{2}/2\right]$. 
For our purposes, we shall be interested in defining a normalization
constant ${\cal N}_{2}$ such that the operator $\phi_{2}\equiv{\cal N}_{2}\left[\frac{\phi^{2}}{2}\right]$
is normalized as follows
\begin{eqnarray}
\Bigr\langle\phi_{2}\left(0,x\right)\phi_{2}\left(0,0\right)\Bigr\rangle\,=\,{\cal N}_{2}^{2}\Bigr\langle\left[\frac{\phi^{2}}{2}\right]\left(0,x\right)\left[\frac{\phi^{2}}{2}\right]\left(0,0\right)\Bigr\rangle & = & \frac{1}{x^{2\Delta_{2}}}\,.
\end{eqnarray}
The explicit value of ${\cal N}_2$
is given in Appendix \ref{app:details_phi2}.

\subsection{A consequence of the time-reversal symmetry}

Let us show that the renomalized composite operator $\left[\bar{\phi}\phi\right]$
is related to $\left[\phi^{2}/2\right]$ by the time reversal
symmetry (\ref{eq:TRS-bare-fields}). 
As a first step, let us consider
the transformation of the monomial $\phi_{0}\bar{\phi}_{0}$:
\begin{eqnarray}
\phi_{0}^{\prime}\left(t,x\right)\bar{\phi}_{0}^{\prime}\left(t,x\right) & = & \phi_{0}\left(-t,x\right)\bar{\phi}_{0}\left(-t,x\right)-\frac{1}{D_{0}}\phi_{0}\left(-t,x\right)\dot{\phi}_{0}\left(-t,x\right)\\
 & = & \phi_{0}\left(-t,x\right)\bar{\phi}_{0}\left(-t,x\right)+\frac{1}{D_{0}}\partial_{t}\left(\frac{\phi_{0}\left(-t,x\right)^{2}}{2}\right)\,.
\end{eqnarray}
The last term on the RHS takes almost the form of the composite operator
$\partial_{t}\left[\phi^{2}/2\right]$. If we multiply the previous
expression by the suitable renormalization factor, i.e., $Z_{22}$,
one finds
\begin{eqnarray}
Z_{D}Z_{22}Z_{\phi}^{1/2}Z_{\bar{\phi}}^{1/2}\phi^{\prime}\left(t,x\right)\bar{\phi}^{\prime}\left(t,x\right) & = & Z_{D}Z_{22}Z_{\phi}^{1/2}Z_{\bar{\phi}}^{1/2}\phi\left(-t,x\right)\bar{\phi}\left(-t,x\right)+\frac{1}{D}\partial_{t}\left[\frac{\phi\left(-t,x\right)^{2}}{2}\right]\,.
\end{eqnarray}

Thus, if we consider the TRS in a correlation function one obtains
\begin{eqnarray}
& &
\Bigr\langle Z_{D}Z_{22}Z_{\phi}^{1/2}Z_{\bar{\phi}}^{1/2}\phi\left(t,x\right)\bar{\phi}\left(t,x\right)\Psi\left(\left\{ t_{i},x_{i}\right\} \right)\Bigr\rangle =  \label{eq:phi2-phibphi-TRS} \\
& &  
\Bigr\langle Z_{D}Z_{22}Z_{\phi}^{1/2}Z_{\bar{\phi}}^{1/2}\phi\left(-t,x\right)\bar{\phi}\left(-t,x\right)\Psi_{{\rm TRS}}\left(\left\{ t_{i},x_{i}\right\} \right)
\Bigr\rangle
+\Bigr\langle\frac{1}{D}\partial_{t}\left[\frac{\phi\left(-t,x\right)^{2}}{2}\right]\Psi_{{\rm TRS}}\left(\left\{ t_{i},x_{i}\right\} \right)\Bigr\rangle\,,
\nonumber
\end{eqnarray}
where $\Psi\left(\left\{ t_{i},x_{i}\right\} \right)$ denotes a bunch
of $\phi$- and $\bar{\phi}$-fields and $\Psi_{{\rm TRS}}\left(\left\{ t_{i},x_{i}\right\} \right)$
its transformation under TRS. The last correlation in (\ref{eq:phi2-phibphi-TRS})
is finite since it involves renormalized fields and a renormalized
composite operator. The other two correlation functions involve the
same operator, $Z_{D}Z_{22}Z_{\phi}^{1/2}Z_{\bar{\phi}}^{1/2}\phi\bar{\phi}$,
evaluated at different times and inserted in different correlation
functions. By assuming the absence of non-trivial systematic cancellations
between these two correlation functions it follows that also the first
two correlations appearing in (\ref{eq:phi2-phibphi-TRS}) are each finite on their own. 
In turn, this implies that the renormalization
factor $Z_{D}Z_{22}$ is the additional renormalization required by
$\left[\phi\bar{\phi}\right]$:
\begin{eqnarray}
\left[\phi\bar{\phi}\right] & = & Z_{\phi\bar{\phi}}Z_{\phi}^{1/2}Z_{\bar{\phi}}^{1/2}\phi\left(t,x\right)\bar{\phi}\left(t,x\right)\\
 &  & \mbox{with }Z_{\phi\bar{\phi}}=Z_{D}Z_{22}\,.
\end{eqnarray}
Actually, the assumption of absence of cancellations in (\ref{eq:phi2-phibphi-TRS})
is not necessary. Indeed, it is possible to check that $\left[\phi\bar{\phi}\right]$
cannot mix with any other operator.

At the fixed point, the TRS entails a relation between the scaling
dimension of $\left[\phi^{2}/2\right]$ and $\left[\phi\bar{\phi}\right]$.
Given $Z_{\phi\bar{\phi}}=Z_{D}Z_{22}$, it is straightforward to check
that
\begin{eqnarray}
\gamma_{\phi^{2}}\,=\,-Z_{22}^{-1}\dot{Z}_{22} & = & 2-z+\gamma_{\phi\bar{\phi},\phi\bar{\phi}}\,,
\end{eqnarray}
where $\gamma_{\phi\bar{\phi},\phi\bar{\phi}}\equiv-Z_{\phi\bar{\phi}}^{-1}\dot{Z}_{\phi\bar{\phi}}$.
Hence the following relation between the scaling dimensions holds:
\begin{eqnarray}
\Delta_{\phi\bar{\phi}} & = & d_{\phi\bar{\phi}}+\gamma_{\phi\bar{\phi},\phi\bar{\phi}}\nonumber \\
 & = & d_{\phi^{2}}+\gamma_{\phi^{2}}+z\,=\,\Delta_{\phi^{2}/2}+z\,.\label{eq:rel_Delta_phi2-Delta_phibphi}
\end{eqnarray}
At order $O\left(g\right)$ it is straightforward to check that $\gamma_{\phi\bar{\phi},\phi\bar{\phi}}=\gamma_{\phi^{2}}+O\left(g^{2}\right)$
with $\gamma_{D}=O\left(g^{2}\right)$. The relation (\ref{eq:rel_Delta_phi2-Delta_phibphi})
has also been discussed in \cite{Calabrese:2004kn}, to which we refer for a more
detailed discussion regarding the TRS.

\section{Gaussian operator product expansion \label{sec:Gaussian-OPE}}

In this section, we consider the OPE for the Gaussian theory. As the
theory is Gaussian one can check the expected results in different
ways. We found it instructive to calculate the OPE expansion by developing
a sort of normal ordering prescription, thus mimicking standard quantum
field theory in which the Gaussian OPE can be derived from the Wick's theorem \cite{Bonneau:2009}.
In particular, we shall see that the following
statements hold:
\begin{eqnarray}
\phi\left(t,x\right)\phi\left(0,0\right) & = & G_{\phi\phi}\left(t,x\right)+\left[\phi^{2}\right]+\mbox{higher derivatives} \label{eq:Gaussian_OPE_phiXphi-1}\\
\phi\left(t,x\right)\bar{\phi}\left(0,0\right) & = & G_{\phi\bar{\phi}}\left(t,x\right)+\left[\phi\bar{\phi}\right]+\mbox{higher derivatives}
\label{eq:Gaussian_OPE_phiXphib-1} \,,
\end{eqnarray}
where $\left[\phi^{2}\right]$ and $\left[\phi\bar{\phi}\right]$
are defined in terms of the normal ordering prescription, 
which we shall introduce later on.
The perturbative corrections to (\ref{eq:Gaussian_OPE_phiXphi-1}) and (\ref{eq:Gaussian_OPE_phiXphib-1}) will be calculated in section \ref{sec:corrections-to-OPE}.
Let us note that in the interacting theory the normal ordering prescription will no longer be sufficient to define a composite operator and we will consider the composite operators introduced in section \ref{sec:Composite-operators}.
In this section, however, we find it useful to derive the Gaussian OPE by adopting the normal ordering prescription.

The RHS of equations (\ref{eq:Gaussian_OPE_phiXphi-1}) and (\ref{eq:Gaussian_OPE_phiXphib-1}) define the OPE coefficients
associated with $\phi \times \phi$ and  $\phi \times \bar{\phi}$. 
For instance, one recognizes that the most singular coefficient of the $\phi\times\phi $ OPE is associated with the identity operator, which we shall denote by
$C_{\phi\phi1}$ with $C_{\phi\phi1}= G_{\phi\phi}\left(t,x\right)$ in the present case.
More generally, we shall denote the coefficient associated with an operator $O$ in the product $O_1\times O_2$ by $C_{O_1O_2O}$.
Finally, let us note the absence in the $\phi\times\bar{\phi}$
OPE of the operator $\left[\phi^{2}\right]$ at the Gaussian level.

\subsection{Gaussian Fokker-Planck Hamiltonian \label{subsec:Gaussian-Fokker-Planck-Hamiltonian}}

We regard the Fokker-Planck equation (\ref{eq:functional_FP_eq_and_FP_Hamiltonian})
as a sort of (functional) Schroedinger equation with an imaginary
time. This allows us to work in the Schroedinger picture and eventually
move to the Heisenberg picture. We refer the reader to \cite{Hatfield:1992rz}
for an overview of the Schroedinger picture of QFT and to \cite{Sakita:1983eq}
for its application to the Fokker-Planck equation in connection to the stochastic
quantization program. 
As a first step, we introduce the following analog
of the position and momentum operators:
\begin{eqnarray}
\hat{\phi}\left(x\right) & = & \phi\left(x\right)\\
\hat{\pi}\left(x\right) & = & \frac{1}{i}\frac{\delta}{\delta\phi\left(x\right)}\,.
\end{eqnarray}
By using the above operators, the Fokker-Planck equation for the free theory can be written as follows:
\begin{eqnarray}
\partial_{t}P\left[\phi\right]\left(t\right) & = & D\Biggr\{\int_{x}i\hat{\pi}\left(x\right)\left(-\partial_{x}^{2}+m^{2}\right)\hat{\phi}\left(x\right)-\int_{x}\hat{\pi}\left(x\right)^{2}\Biggr\} P\left[\phi\right]\left(t\right)\,.
\end{eqnarray}
Note that the first term is non-Hermitian. Therefore, the ``Hamiltonian''
on the RHS is not Hermitian as well.\footnote{Nevertheless, it is interesting to note that $H_{{\rm FP}}$ appears
to be ``PT-symmetric'' in the sense that it is invariant under $\pi\rightarrow \pi$,
$\phi\rightarrow-\phi$, and $i\rightarrow-i$.} 
Next, we introduce the following operators:
\begin{eqnarray}
\hat{a}\left(p\right) & \equiv & \beta\Delta_{p}\left(\hat{\phi}\left(p\right)+i\frac{1}{\Delta_{p}}\hat{\pi}\left(p\right)\right)\label{eq:def-a_p}\\
\hat{b}\left(p\right) & \equiv & \frac{1}{i\beta\Delta_{p}}\hat{\pi}\left(-p\right)\,,\label{eq:def-b_p}
\end{eqnarray}
where $\beta$ is an arbitrary parameter and $\Delta_{p}\equiv\left(p^{2}+m^{2}\right)$.
The operators $\hat{a}$ and $\hat{b}$ satisfy the following algebra:
\begin{eqnarray}
\left[\hat{a}\left(p\right),\hat{b}\left(q\right)\right] & = & \left(2\pi\right)^{d}\delta\left(p-q\right)\,,
\end{eqnarray}
which is the same as that of creation-annihilation operators (see, e.g., \cite{Peskin:1995ev}) but with $b$ being replaced by $a^{\dagger}$.

A straightforward calculation shows then that the Fokker-Planck Hamiltonian
takes the following form
\begin{eqnarray}
H_{{\rm FP}} & = & D\int_{k}\Delta_{k}\left(\hat{b}\left(k\right)\hat{a}\left(k\right)\right)\,=\,\int_{k}\omega_{k}\left(\hat{b}\left(k\right)\hat{a}\left(k\right)\right)\,, \label{eq:def_H_FP_normal_ordered}
\end{eqnarray}
with $\omega_{k}\equiv D\Delta_{k}$. Again, let us note that since
both $\hat{a}$ and $\hat{b}$ are not Hermitian also $H_{{\rm FP}}$
is not.

\subsection{Normal ordering \label{subsec:Normal-ordering}}

One can re-express the original fields $\phi$ and $\bar{\phi}$ in
terms of the creation and annihilation operators (\ref{eq:def-a_p}--\ref{eq:def-b_p}).
The Heisenberg operators are defined by
\begin{eqnarray}
\phi\left(t,x\right) & \equiv & e^{H_{{\rm FP}}t}\phi\left(x\right)e^{-H_{{\rm FP}}t}\\
\bar{\phi}\left(t,x\right) & \equiv & e^{H_{{\rm FP}}t}\bar{\phi}\left(x\right)e^{-H_{{\rm FP}}t}\,.
\end{eqnarray}
Finally, let us introduce the normal ordering operation, which we
denote by sandwiching any operator in colons, i.e., $:\hat{O}:$.
We define normal ordering as the operation that maps a given operator
$\hat{O}\left(\hat{a},\hat{b}\right)$ to a new operator obtained from $\hat{O}$ by moving the creation and annihilation operators
so that the $\hat{b}$-operators are on left of the $\hat{a}$-operators.
For example, the Hamiltonian (\ref{eq:def_H_FP_normal_ordered}) is normal ordered.

It is then possible to follow a textbook presentation of the Wick's theorem and check that the following statements holds:
\begin{eqnarray}
T\phi\left(t_{2},x_{2}\right)\phi\left(t_{1},x_{1}\right) & = & G_{\phi\phi}\left(t_{2}-t_{1},x_{2}-x_{1}\right)+:\phi\left(t_{1},x_{1}\right)\phi\left(t_{2},x_{2}\right):\\
T\phi\left(t_{2},x_{2}\right)\bar{\phi}\left(t_{1},x_{1}\right) & = & G_{\bar{\phi}\phi}\left(t_{2}-t_{1},x_{2}-x_{1}\right)+:\phi\left(t_{2},x_{2}\right)\bar{\phi}\left(t_{1},x_{1}\right):\,,
\end{eqnarray}
where the $T$ on the left hand side denotes the time ordering.

At this point, the OPE amounts to the Taylor expansion of the normal
ordering statement:
\begin{equation}
\boxed{
T\phi\left(t_{2},x_{2}\right)\phi\left(t_{1},x_{1}\right)  =
G_{\phi\phi}\left(t_{2}-t_{1},x_{2}-x_{1}\right)+:\phi\left(t_c,x_c\right)^{2}:+\cdots
}
\end{equation}
and
\begin{equation}
\boxed{
T\phi\left(t_{2},x_{2}\right)\bar{\phi}\left(t_{1},x_{1}\right) = 
G_{\bar{\phi}\phi}\left(t_{2}-t_{1},x_{2}-x_{1}\right)+:\phi\left(t_c,x_c\right)\bar{\phi}\left(t_c,x_c\right):+\cdots\,,}
\end{equation}
where $t_c \equiv \left(t_1+t_2\right)/2$ and $x_c \equiv \left(x_1+x_2\right)/2$.
It should be noted that, strictly speaking, the normal ordering represents
one possible regularization for the composite operators of the free
theory. In particular, the expectation value on any normal ordered
operator vanishes by construction. However, this is just a possible
choice and other choices, such as dim-reg discussed in section \ref{subsec:The-composite-operator-phi2},
lead to different values of the VEV. Clearly, also the associated
OPE coefficients are scheme dependent in general. 
However, one can obtain universal and scheme independent results,
as we shall check explicitly at the fixed point. 

Finally, let us consider the OPE of $\bar{\phi}\times\bar{\phi}$,
at the Gaussian level one has
\begin{equation}
\boxed{
T\bar{\phi}\left(t_{2},x_{2}\right)\bar{\phi}\left(t_{1},x_{1}\right) =
:\bar{\phi}\left(t_c,x_c\right)^{2}:+\cdots\,.}
\end{equation}
It is also straightforward to check that $C_{\bar{\phi}\bar{\phi}1}=C_{\bar{\phi}\bar{\phi}\left[\phi^{2}\right]}=0$
exactly. 
To see this, first consider the OPE within $0=\langle\bar{\phi}\bar{\phi}\rangle$:
this relation implies that only operators having vanishing VEVs appear in the OPE. 
In particular $C_{\bar{\phi}\bar{\phi}1}=0$. Next, let
us consider the OPE of $\bar{\phi}\times\bar{\phi}$ in $0=\langle\bar{\phi}\bar{\phi}\bar{\phi}\bar{\phi}\rangle$.
Given $C_{\bar{\phi}\bar{\phi}1}=0$, the most singular contribution
comes from $0=\langle C_{\bar{\phi}\bar{\phi}\left[\phi^{2}\right]}\left[\phi^{2}\right]\bar{\phi}\bar{\phi}\rangle$.
Since $\langle\left[\phi^{2}\right]\bar{\phi}\bar{\phi}\rangle\neq0$
at tree level one must have $C_{\bar{\phi}\bar{\phi}\left[\phi^{2}\right]}=0$.

\section{Leading corrections to the OPE \label{sec:corrections-to-OPE}}

In this section we consider the OPE associated with the products $\phi\times\phi$
and $\phi\times\bar{\phi}$ and compute the most singular contributions to the first order in perturbation theory. 
We shall focus on deriving the OPE coefficients at the fixed point.
Let us recall that the OPE applies to the short-distance limit.
In our case, for instance, we define the short-distance limit of $\phi \left(t,x\right) \phi \left(0,0\right)$ as the limit $x\rightarrow 0$ with $t/x^z$ being held fixed.

One may adopt various strategies to compute OPE coefficients: 
for instance, one may look at the three- or four-point function
and consider the short distance limit of two of the operators appearing
in such correlations, see, e.g., \cite{Collins:1984xc}. 
In this section we follow a different approach,
based on the mass expansion of the two-point function, 
which leads to the same result,
as we have checked explicitly. 
On the one hand, this approach allows one to derive the OPE coefficients and at the same time to calculate
the two-point function close to criticality, i.e., for values of $m$
small but non-zero. Indeed, one of the first applications of the OPE
was precisely the prediction of the form of the static two-point function
close to criticality in the regime $m\ll p$ \cite{Brezin:1974zz,Brezin1974b,Caselle:2015csa}.
For a modern use of this idea in the static case see
\cite{Caselle:2015csa,Costagliola:2015ier,Caselle:2016mww}.
Let us mentioned, however, that 
we checked that the results of this section can also be obtained by looking at the three- and four-point functions. 
In this way one has a non-trivial check of the consistency of the computed OPE coefficients.

As we already mentioned, we shall work in the MS scheme. This implies
that the OPE coefficients are analytic in the mass, see 
\cite{Tkachov:1983st,Guida:1995kc}
and references therein. It follows that the non-analytic dependence
of the two-point function on the mass is entirely due to the VEV of
the operators once the OPE has been performed. 
A key point here is
also that the most singular OPE coefficients are multiplied by 
the operators that have the smallest mass dimension. For instance, in
4D one expects $\langle\phi^{2}\rangle\sim m^{2}$ while $\langle\phi^{4}\rangle\sim m^{4}$.
Of course, fluctuations modify the exponents as dictated by the renormalization
group. Let us consider the product $\phi\times\phi$ as an example,
one has
\begin{eqnarray}
\langle\phi\left(t,x\right)\phi\left(0,0\right)\rangle 
& \approx & \left(C_{\phi\phi1}^{\left(0\right)}\left(t,x\right)+m^{2}C_{\phi\phi1}^{\left(1\right)}\left(t,x\right)+\cdots\right)+\left(C_{\phi\phi\left[\frac{\phi^{2}}{2}\right]}^{\left(0\right)}\left(t,x\right)+\cdots\right)\Bigr\langle\left[\frac{\phi^{2}}{2}\right]\Bigr\rangle 
\nonumber \\
& & +o\left(m^{2}\right)\,, \label{eq:2pt-mass-exp}
\end{eqnarray}
where $\langle\left[\phi^{2}\right]\rangle\sim m^{2}$
and the superscript $(n)$ in $C^{(n)}$ denotes the order of the expansion in the mass paramter. 
By expanding
the LHS of (\ref{eq:2pt-mass-exp}) one can calculate the OPE coefficients
since they are the only unknowns after inserting the value calculated
for the VEVs. We remark that the coefficients $C_{\phi\phi1}^{\left(0\right)}\left(t,x\right)$
and $C_{\phi\phi\left[\phi^{2}\right]}^{\left(0\right)}\left(t,x\right)$
evaluated at $g=g_{*}$ are actually the fixed point coeffients, which
is our main interest.

\subsection{OPE expansion of $\langle\phi\left(t,x\right)\phi\left(0,0\right)\rangle$
\label{subsec:OPE-expansion-of-phiphi}}

In section \ref{sec:Composite-operators} we have computed the VEV
of the operator $\left[\phi^{2}/2\right]$ up to order $O\left(g\right)$.
In order to apply the strategy outlined at the beginning of this section,
we must compute the LHS of equation (\ref{eq:2pt-mass-exp}) up to
the same order. At the tree level the two-point function is given
by equation (\ref{eq:Gphiphi-Gaussian}). 
The diagrams contributing to
order $O\left(g\right)$ only involve the mass renormalization. Thus,
it turns out that the LHS of (\ref{eq:2pt-mass-exp}) is given precisely
by (\ref{eq:Gphiphi-Gaussian}) once the substitution
\begin{eqnarray}
m^{2} & \rightarrow & m^{2}+\frac{1}{2}g\frac{m^{2}}{\left(4\pi\right)^{2}}\left(\log\left(\frac{m^{2}}{4\pi\mu^{2}}\right)+\gamma-1\right)\label{eq:mass-corr-Og}
\end{eqnarray}
is made and the expansion to first order in $g$ is performed. (We
work to the leading order in $\epsilon$.)

Let us note that in the present case $\langle\phi\left(t,x\right)\phi\left(0,0\right)\rangle$
is not at the fixed point: we shall set $g=g_{*}$ but keep a non-zero,
albeit small, mass so that we are a bit off criticality. 
By matching each term in the expansions of the LHS and RHS of equation (\ref{eq:2pt-mass-exp}),
one obtains the following results:
\begin{eqnarray}
C_{\phi\phi1}^{\left(0\right)}\left(t,x\right) & = & \frac{2^{d-2}\Gamma\left(\frac{d}{2}-1\right)}{\left(4\pi\right)^{d/2}}\frac{1}{x^{d-2}}\Biggr\{1-\frac{\Gamma\left(\frac{d}{2}-1,\frac{x^{2}}{4D|t|}\right)}{\Gamma\left(\frac{d}{2}-1\right)}\Biggr\}+O\left(\epsilon^{2}\right)\\
C_{\phi\phi\left[\frac{\phi^{2}}{2}\right]}^{\left(0\right)}\left(t,x\right) & = & 2+g_{*}\frac{1}{16\pi^{2}}\Biggr\{\log\left(\pi x^{2}\mu\right)+\gamma+\Gamma\left(0,\frac{x^{2}}{4D|t|}\right)\Biggr\}+O\left(\epsilon^{2}\right)\,,
\end{eqnarray}
where $\Gamma\left(a,b\right)\equiv\int_{b}^{\infty}dt\,e^{-t}t^{a-1}$.

Let us remark that the same result for the OPE coefficients can be obtained by looking at the four-point function, similarly to what is done in \cite{Collins:1984xc}.

\subsection{OPE expansion of $\langle\phi\left(t,x\right)\bar{\phi}\left(0,0\right)\rangle$
\label{subsec:OPE-expansion-of-phiphib}}

We consider the OPE expansion of $\langle\phi\left(t,x\right)\bar{\phi}\left(0,0\right)\rangle$,
i.e, the response function. 
In analogy with the reasoning of section \ref{subsec:OPE-expansion-of-phiphi}, 
we first consider the mass expansion
of the response function to order $O\left(g\right)$. 
To do so, we consider the Gaussian expression, equation (\ref{eq:eq:Gphiphib-Gaussian}).
The order $O\left(g\right)$ is reproduced by implementing the substitution
(\ref{eq:mass-corr-Og}) and expanding in the coupling constant. Then
we consider
\begin{eqnarray}
\langle\phi\left(t,x\right)\bar{\phi}\left(0,0\right)\rangle & \approx & \left(C_{\phi\bar{\phi}1}^{\left(0\right)}\left(t,x\right)+m^{2}C_{\phi\bar{\phi}1}^{\left(1\right)}\left(t,x\right)+\cdots\right)+\left(C_{\phi\bar{\phi}\left[\frac{\phi^{2}}{2}\right]}^{\left(0\right)}\left(t,x\right)+\cdots\right)\Bigr\langle\left[\frac{\phi^{2}}{2}\right]\Bigr\rangle \nonumber \\
&& +o\left(m^{2}\right)\,.
\end{eqnarray}
The LHS is proportional to $\theta\left(t\right)$ and so are the
OPE coefficients on the RHS. 

By matching the terms on the LHS and RHS, we deduce
\begin{eqnarray}
C_{\phi\bar{\phi}1}^{\left(0\right)}\left(t,x\right) & = & \frac{\theta\left(t\right)}{\left(4\pi\right)^{d/2}}\frac{e^{-\frac{x^{2}}{4Dt}}}{\left(Dt\right)^{d/2}}+O\left(\epsilon^{2}\right)\\
C_{\phi\bar{\phi}\left[\frac{\phi^{2}}{2}\right]}^{\left(0\right)}\left(t,x\right) & = & -g_{*}\frac{\theta\left(t\right)}{\left(4\pi\right)^{2}}\frac{e^{-\frac{x^{2}}{4Dt}}}{\left(Dt\right)}+O\left(\epsilon^{2}\right)\,,
\end{eqnarray}
where we did not expand $C_{\phi\bar{\phi}1}^{\left(0\right)}$ in
$\epsilon$ while the coefficient $C_{\phi\bar{\phi}\left[\frac{\phi^{2}}{2}\right]}^{\left(0\right)}$
is already expressed to $O\left(\epsilon\right)$.

\subsection{Relation to the fluctuation-dissipation theorem}

The time-reversal symmetry (\ref{eq:action_model_A-1}) implies exact
relations among various correlation functions. It would be interesting
to study in detail the consequence of the symmetry (\ref{eq:action_model_A-1})
for the OPE. In this section we limit ourselves to the following consideration.
Let us recall the fluctuation-dissipation theorem:
\begin{eqnarray}
\langle\phi\left(t,x\right)\bar{\phi}\left(0,0\right)\rangle & = & -\frac{1}{D}\partial_{t}\langle\phi\left(t,x\right)\phi\left(0,0\right)\rangle\,,\label{eq:FDT-for-2pt-function}
\end{eqnarray}
where $t>0$ and we work in the MS-scheme. We can expand via the OPE
each of the correlation functions appearing in (\ref{eq:FDT-for-2pt-function}).
One obtains
\begin{eqnarray}
\Bigr\langle C_{\phi\bar{\phi}1}\left(t,x\right)+C_{\phi\bar{\phi}\left[\frac{\phi^{2}}{2}\right]}\left(t,x\right)\left[\frac{\phi^{2}}{2}\right]+\cdots\Bigr\rangle & = & -\frac{1}{D}\partial_{t}\Bigr\langle C_{\phi\phi1}\left(t,x\right)+C_{\phi\phi\left[\frac{\phi^{2}}{2}\right]}\left(t,x\right)\left[\frac{\phi^{2}}{2}\right]+\cdots\Bigr\rangle\,.
\nonumber \\
& &
\end{eqnarray}
Again, we may consider the above relation off-criticality and consider
the $m^{2}$-expansion. The leading term in this expansion is determined
by the OPE coefficients associated with the identity, which,
in the small mass limit,
must satisfy 
\begin{equation}
\boxed{
C_{\phi\bar{\phi}1}\left(t,x\right) =
-\frac{1}{D}\partial_{t}C_{\phi\phi1}\left(t,x\right)\,.
}
\end{equation}
Moreover, since the non-analytic dependence on the mass is captured
by the VEV of composite operators, one also obtains
\begin{equation}
\boxed{
C_{\phi\bar{\phi}\left[\frac{\phi^{2}}{2}\right]}\left(t,x\right)  =
-\frac{1}{D}\partial_{t}C_{\phi\phi\left[\frac{\phi^{2}}{2}\right]}\left(t,x\right)\,
}
\end{equation}
in the small mass limit.
It is straightforward to check that the coefficients calculated in
sections \ref{subsec:OPE-expansion-of-phiphi} and \ref{subsec:OPE-expansion-of-phiphib}
satisfy the above non-perturbative relations.

\section{Universal quantities from the OPE \label{sec:Universal-quantities-from-OPE}}

In the previous section we obtained certain OPE coefficients. Now
we shall see that they entail universal information after implementing
suitable normalization conditions. The arguments used in this section
follow the line of reasoning detailed in \cite{Amit:1984ms}, 
to which we refer for a review of universality in the context of the perturbative
RG and its application to the statistical field theory approach for
critical phenomena.

\subsection{Universality in the static limit \label{subsec:Universality-in-the-static-lim}}

In critical systems the most calculated universal quantities
are, by far, the critical exponents, which turn out to display universality in a rather direct way. 
There are, however, further quantities that
display universality. In particular, the so-called universal amplitude ratios and the universal form of the equation of state have been studied
\cite{Brezin:1972fc,Brezin:1972fb,BREZIN1974285,Bervillier:1976zz}. 
In order to define such quantities,
the implementation of some convention is required.
For instance, in the case of the equation of state two normalizations must be imposed \cite{Amit:1984ms}.
In the case of of the universal equation of state, initially
one calculates a certain expression, which must
be consistent with the associated Callan-Symanzik (CS) equation.
The analysis of the CS equation allows one to pinpoint where the
non-universal contributions come from and it turns out that a scaling
variable\footnote{It can be a field or a coupling, such as the mass in the case of the
universal equation of state.} is typically multiplied by a factor that depends on the integrated
trajectory along the RG flow rather than only on the fixed point itself.
The idea is then to define one or more new variables by imposing certain
conventions in such a way that this non-universal factor is ``eaten up''
by the new variables and one is left with a universal result.

In order to make contact with the OPE coefficients, 
let us start by
considering the static case. In many cases of interest, the static
limit is described not just by a scale-invariant theory but by a CFT.
In the CFT framework, once a normalization is fixed for the operators,
typically $\langle O\left(x\right)O\left(0\right)\rangle=1/x^{2\Delta_{O}}$,
the OPE coefficients take the form $c_{ijk}/x^{\Delta_{i}+\Delta_{j}-\Delta_{k}}$
where the number $c_{ijk}$ is as universal as the scaling dimensions
$\Delta_{i}$ since both are determined solely on the basis of symmetry
and field content. 
Let us provide a brief argument from the field theory perspective
via the following example: we consider the scaling
field $\phi_{1}\equiv{\cal N}_{1}\phi$ and $\phi_{2}\equiv{\cal N}_{2}\left[\phi^{2}/2\right]$.\footnote{The elementary field $\phi$ is a scaling operator. 
For a composite operator, one may have to take a suitable combination of composite
operators.} The normalizations ${\cal N}_{i}$ are determined by imposing $\langle\phi_{i}\left(x\right)\phi_{j}\left(0\right)\rangle=\delta_{ij}/x^{2\Delta_{i}}$.
Next, we consider the ratio
\begin{eqnarray}
\frac{\langle\phi\left(p_{1}\right)\phi\left(p_{2}\right)\left[\frac{\phi^{2}}{2}\right]\left(-p_{1}-p_{2}\right)\rangle}{\langle\phi\left(p_{1}-p_{2}\right)\phi\left(-p_{1}+p_{2}\right)\rangle\langle\left[\frac{\phi^{2}}{2}\right]\left(p_{1}+p_{2}\right)\left[\frac{\phi^{2}}{2}\right]\left(-p_{1}-p_{2}\right)\rangle^{1/2}}\,.
\end{eqnarray}
It is straightforward to check
that this ratio can be rewritten as follows
\begin{eqnarray}
\frac{\langle\phi_{1}\left(p_{1}\right)\phi_{1}\left(p_{2}\right)\phi_{2}\left(-p_{1}-p_{2}\right)\rangle}{\langle\phi_{1}\left(p_{1}-p_{2}\right)\phi_{1}\left(-p_{1}+p_{2}\right)\rangle\langle\phi_{2}\left(p_{1}+p_{2}\right)\phi_{2}\left(-p_{1}-p_{2}\right)\rangle^{1/2}}\,.
\end{eqnarray}
The two-point function in the denominator are just the Fourier transform
of our normalization condition $\langle\phi_{1}\left(p\right)\phi_{1}\left(-p\right)\rangle=c_{1}p^{\alpha_{1}}$
and $\langle\phi_{2}\left(p\right)\phi_{2}\left(-p\right)\rangle=c_{2}p^{\alpha_{2}}$
calculated for some given momenta. The striking feature of this ratio
is that the associated CS equation does not depend on the wave functions
renormalizations of $\phi$ and of $\left[\phi^{2}\right]$ implying
that the following equation holds at zero mass:
\begin{eqnarray}
\left(\mu\partial_{\mu}+\beta\partial_{g}\right)\frac{\langle\phi_{1}\left(p_{1}\right)\phi_{1}\left(p_{2}\right)\phi_{2}\left(-p_{1}-p_{2}\right)\rangle}{\langle\phi_{1}\left(p_{1}-p_{2}\right)\phi_{1}\left(-p_{1}+p_{2}\right)\rangle\langle\phi_{2}\left(p_{1}+p_{2}\right)\phi_{2}\left(-p_{1}-p_{2}\right)\rangle^{1/2}} & = & 0\,.
\end{eqnarray}
By using the arguments detailed in \cite{Amit:1984ms}, one can see that the above ratio is universal. 
Thus, by applying the OPE to the three-point function above 
(i.e., taking a suitable large momentum limit) 
it is straightforward to conclude that the coefficient $c_{112}$ is universal too.

\subsection{Universality in the dynamical case}

\subsubsection{General discussion}

Let us consider a two-point function $C\left(t,x\right)$, such as
$\langle\phi\phi\rangle$ or $\langle\phi\bar{\phi}\rangle$. At
a fixed point scale invariance implies that
\begin{eqnarray}
C\left(t,x\right) & = & \frac{1}{x^{\Delta}}F\left(c_{1}\frac{t}{x^{z}}\right)\,,
\end{eqnarray}
where $c_{1}$ is a non-universal constant. If $F\left(0\right)=c_{2}$
one can rewrite
\begin{eqnarray}
C\left(t,x\right) & = & \frac{c_{2}}{x^{\Delta}}f\left(c_{1}\frac{t}{x^{z}}\right)\,,\label{eq:2pt-funct-generic-universality-1}
\end{eqnarray}
where $f\left(0\right)=1$ and $c_{2}$ is a non-universal amplitude.
The form of the function $f$ is universal and is fully specified
once a further condition is imposed. This is needed since one could
equally well consider a constant $\tilde{c}_{1}=\alpha c_{1}$, with
$\alpha$ being a numerical factor. Therefore, we introduce the scaling
variable $y={\cal N}_{y}t/x^{z}$ and write
\begin{eqnarray}
C\left(t,x\right) & = & \frac{c_{2}}{x^{\Delta}}f\left(c_{1}{\cal N}_{y}^{-1}y\right)\,\equiv\,\frac{c_{2}}{x^{\Delta}}f_{y}\left(y\right)\,,
\end{eqnarray}
where ${\cal N}_{y}$ is determined by imposing a suitable condition,
such as
\begin{eqnarray}
f_{y}\left(y=1\right) & = & {\rm const}.
\end{eqnarray}

It is important to note that the scaling variable $y$ can be used
in all two-point functions since they depend on the combination $c_{1}t/x^{z}$,
where the non-universal constant $c_{1}$ is related to the RG running
of the coupling $D$.

Let us now move to consider more specifically the OPE coefficients.
At the fixed point, scale invariance implies that a OPE coefficient takes the form
\begin{eqnarray}
C_{ijk}\left(t,x\right) & = & \frac{1}{x^{\Delta_{i}+\Delta_{j}-\Delta_{k}}}F_{ijk}\left(c_{1}\frac{t}{x^{z}}\right)\,.\label{eq:OPE-coeff-at-FP-form-1}
\end{eqnarray}
In the static limit, we recover the equilibrium case and the discussion
of section \ref{subsec:Universality-in-the-static-lim} applies. Thus,
one can still normalize the operators as discussed in section \ref{subsec:Universality-in-the-static-lim}.
For example, we define the scaling field $\phi_{1}\equiv{\cal N}_{1}\phi$,
where ${\cal N}_{1}$ is determined by
\begin{eqnarray}
\langle\phi_{1}\left(0,x\right)\phi_{1}\left(0,0\right)\rangle & = & \frac{1}{x^{2\Delta_{\phi}}}\,.
\end{eqnarray}
This normalization choice allows one to get rid of non-universal overall
factors like $c_{2}$ in (\ref{eq:2pt-funct-generic-universality-1}).
However, even after this normalization, an OPE coeffient does not yet display universality away from the static limit due to the presence
of $c_{1}$ in (\ref{eq:OPE-coeff-at-FP-form-1}). Nevertheless, it
is now straightforward to obtain a universal form of the OPE coefficient
by moving to the normalized scaling variable $y$. 
Once expressed in this variable, 
the OPE coefficient of a normalized scaling operator
will take the form
\begin{eqnarray}
C_{ijk}\left(t,x\right) & = & \frac{1}{x^{\Delta_{i}+\Delta_{j}-\Delta_{k}}}c_{ijk}\left(y\right)\,,
\end{eqnarray}
where $c_{ijk}\left(y\right)$ is a universal scaling function.

\subsubsection{Application to our results \label{subsec:Application-to-our-operators}}

We must first proceed to define suitably normalized scaling fields.
We begin by considering $\phi_{1}\equiv{\cal N}_{1}\phi$ and imposing
\begin{eqnarray}
\langle\phi_{1}\left(0,x\right)\phi_{1}\left(0,0\right)\rangle & = & \frac{1}{x^{2\Delta_{\phi}}}\,,
\end{eqnarray}
with $\Delta_{\phi}=\frac{d-2}{2}$. Within our approximations, this
implies
\begin{eqnarray}
{\cal N}_{1} & = & \left(\frac{2^{d-2}\Gamma\left(\frac{d}{2}-1\right)}{\left(4\pi\right)^{d/2}}\right)^{-1/2}\,.
\end{eqnarray}
We then have
\begin{eqnarray}
\langle\phi_{1}\left(t,x\right)\phi_{1}\left(0,0\right)\rangle & = & \frac{1}{x^{2\Delta_{\phi}}}\Biggr\{1-\frac{\Gamma\left(\frac{d}{2}-1,\frac{x^{2}}{4D|t|}\right)}{\Gamma\left(\frac{d}{2}-1\right)}\Biggr\}\,.
\end{eqnarray}
The above two-point function is clearly non-universal as it depends
on the non-universal coupling $D$. We introduce the scaling variable
$y\equiv{\cal N}_{y} t/x^{z}$, where ${\cal N}_{y}$ is determined
by
\begin{eqnarray}
x^{2\Delta_{\phi}}\langle\phi_{1}\left(t,x\right)\phi_{1}\left(0,0\right)\rangle\Bigr|_{y=1} & = & 1-\frac{\Gamma\left(\frac{d}{2}-1,1\right)}{\Gamma\left(\frac{d}{2}-1\right)}\,.
\end{eqnarray}
This choice is made for pure convenience and others are possible.
We obtain ${\cal N}_{y}=4D$ and rewrite
\begin{eqnarray}
\langle\phi_{1}\left(t,x\right)\phi_{1}\left(0,0\right)\rangle & = & \frac{1}{x^{2\Delta_{\phi}}}g\left(y\right)\mbox{ with }g\left(y\right)=1-\frac{\Gamma\left(\frac{d}{2}-1,\frac{1}{|y|}\right)}{\Gamma\left(\frac{d}{2}-1\right)}\,,
\end{eqnarray}
where the function $g\left(y\right)$ is universal.

Next, we wish to normalize the scaling field $\bar{\phi}_{1}\equiv\bar{{\cal N}}_{1}\phi$
by imposing a suitable normalization condition. Let us note that we
cannot use the two-point $\langle\bar{\phi}_{1}\bar{\phi}_{1}\rangle$
since it vanishes. Therefore, we consider the response function $\langle\phi_{1}\left(t,x\right)\bar{\phi}_{1}\left(0,0\right)\rangle$
for $t>0$, so that it does not vanish due to the presence of $\theta\left(t\right)$.
In particular we impose
\begin{eqnarray}
\langle\phi_{1}\left(t>0,x\right)\bar{\phi}_{1}\left(0,0\right)\rangle\Bigr|_{y=1} & = & \frac{1}{x^{d}}\,,
\end{eqnarray}
which implies $\bar{{\cal N}}_{1}=\frac{\pi^{d/2}e}{{\cal N}_{1}}$.

We are now ready to apply these normalizations to our results. Let
us consider the OPE of the product $\phi_{1}\times\phi_{1}$:
\begin{eqnarray}
\phi_{1}\times\phi_{1} & = & {\cal N}_{1}^{2}\phi\times\phi\,=\,{\cal N}_{1}^{2}C_{\phi\phi1}+{\cal N}_{1}^{2}C_{\phi\phi\left[\frac{\phi^{2}}{2}\right]}\left[\frac{\phi^{2}}{2}\right]+\cdots\\
 & = & {\cal N}_{1}^{2}C_{\phi\phi1}+{\cal N}_{1}^{2}C_{\phi\phi\left[\frac{\phi^{2}}{2}\right]}{\cal N}_{2}^{-1}\phi_{2}+\cdots\\
 & = & C_{\phi_{1}\phi_{1}1}+C_{\phi_{1}\phi_{1}\phi_{2}}\phi_{2}+\cdots\,.
\end{eqnarray}
Thus, we obtain
\begin{subequations}
\begin{empheq}[box=\widefbox]{align}
  C_{\phi_{1}\phi_{1}1}\left(t,x\right) & =  \frac{1}{x^{2\Delta_{\phi_1}}}\left\{ 1-\frac{\Gamma\left(\frac{d}{2}-1,\frac{1}{|y|}\right)}{\Gamma\left(\frac{d}{2}-1\right)}\right\} +O\left(\epsilon^{2}\right)\\
C_{\phi_{1}\phi_{1}\phi_{2}}\left(t,x\right) & =  \frac{1}{x^{2\Delta_{\phi_{1}}-\Delta_{\phi_{2}}}}\sqrt{2}\left(1-\frac{\epsilon}{6}+\frac{\epsilon}{6}\Gamma\left(0,\frac{1}{|y|}\right)\right)+O\left(\epsilon^{2}\right)\,.
\end{empheq}
\end{subequations}
In the present approximation $C_{\phi_{1}\phi_{1}1}\left(t,x\right)$
is fixed by the Gaussian two-point function together with the chosen
normalization conventions. The coefficient $C_{\phi_{1}\phi_{1}\phi_{2}}\left(t,x\right)$
is more interesting: in the static limit, i.e., $y\rightarrow0$,
we recover the $\epsilon$-correction known from CFT techniques 
\cite{Gaiotto:2013nva,Gopakumar:2016wkt,Gopakumar:2016cpb,Dey2017a,Gopakumar:2018xqi}.
In the dynamical case, this universal coefficient is promoted to be
a function of the scaling variable $y$, which begins to be non-trivial
from $O\left(\epsilon\right)$.
Let us note that we expect the equal-time limit of the OPE coefficient to coincide with the associated static coefficients.
In this limit, non-perturbative values have been obtained by a variety of methods, 
such as conformal bootstrap \cite{Kos:2016ysd}, Monte Carlo simulations \cite{Caselle:2015csa,Costagliola:2015ier,Caselle:2016mww}, and functional renormalization group \cite{Rose:2021zdk}.
We refer to \cite{Rose:2021zdk} for a comparison of numerical results of the various approaches.

In full analogy, we obtain the following results for the product $\phi_{1}\times\bar{\phi}_{1}$:
\begin{subequations}
\begin{empheq}[box=\widefbox]{align}
  C_{\phi_{1}\bar{\phi}_{1}1}\left(t,x\right) & =
\frac{\theta(y)}{x^{\Delta_{\phi_{1}}+\Delta_{\bar{\phi}_{1}}}}\frac{4e^{-1/y}y^{-d/2}}{\Gamma\left(\frac{d-2}{2}\right)}+O\left(\epsilon^{2}\right)\\
C_{\phi_{1}\bar{\phi}_{1}\phi_{2}}\left(t,x\right) & =
\frac{\theta(y)}{x^{\Delta_{\phi_{1}}+\Delta_{\bar{\phi}_{1}}-\Delta_{\phi_{2}}}}\left(-\epsilon\frac{e^{1-\frac{1}{y}}y^{-1}}{3\sqrt{2}}\right)+O\left(\epsilon^{2}\right)\,.
\end{empheq}
\end{subequations}

\section{Extension to $O\left(N\right)$ models \label{sec:Extention-to-ON}}

In this section we generalize the previous results to the critical
dynamics associated with $O\left(N\right)$ models. We consider the action
\begin{eqnarray}
S & = & \int_{t,x}\bar{\phi}_{0}^{m}\left(t,x\right)\Biggr\{\partial_{t}\phi_{0}^{m}\left(t,x\right)+D_{0}\left(-\partial^{2}+m_{0}^{2}+\frac{g_{0}}{3!}\phi_{0}^{n}\left(t,x\right)\phi_{0}^{n}\left(t,x\right)\right)\phi_{0}^{m}\left(t,x\right)\\
 &  & -D_{0}\bar{\phi}_{0}^{m}\left(t,x\right)\bar{\phi}_{0}^{m}\left(t,x\right)\Biggr\}\,.
\end{eqnarray}
In $d=4-\epsilon$ at one-loop, the fixed point value of the interaction
coupling is given by $\epsilon\frac{3}{\left(N+8\right)}\left(4\pi\right)^{2}$
while the mass anomalous dimension is modified by a factor $\frac{N+2}{3}$
with respect to the $N=1$ case.

\subsection{OPE at the Gaussian level \label{subsec:OPE-Gauss_ON}}

It is instructive to inspect the OPE of two fields at the Gaussian
level. Let us define the composite operators as normal ordered operators,
in full analogy with section \ref{subsec:Normal-ordering}. One obtains
\begin{eqnarray}
\phi^{a}\times\phi^{b} & = & G_{ab}+2\left[\frac{\phi^{a}\phi^{b}}{2}\right]+\cdots\,.\label{eq:OPE-Gaussian-ON-1}
\end{eqnarray}
For later purposes it is convenient to consider the following combinations
of operators of dimension $d-2$:
\begin{eqnarray}
\left[\frac{\phi^{a}\phi^{b}}{2}\right]_{T} & \equiv & \left[\frac{\phi^{a}\phi^{b}}{2}-\frac{\delta^{ab}}{N}\frac{\phi^{m}\phi^{m}}{2}\right]\\
\left[\frac{\phi^{a}\phi^{b}}{2}\right]_{S} & \equiv & \left[\frac{\delta^{ab}}{N}\frac{\phi^{m}\phi^{m}}{2}\right]\,,
\end{eqnarray}
where the Einstein summation convention is used. By using the above
combination we can rewrite equation (\ref{eq:OPE-Gaussian-ON-1})
as
\begin{eqnarray}
\phi^{a}\times\phi^{b} & = & G_{ab}+2\left[\frac{\phi^{a}\phi^{b}}{2}\right]_{T}+2\left[\frac{\phi^{a}\phi^{b}}{2}\right]_{S}+\cdots\,.
\end{eqnarray}
At this stage the introduction of the ``transverse'' and ``longitudinal''
composite operators may seem arbitrary. However, as we shall see in
the next section, they have different scaling dimensions and also their
role in the OPE is different. This difference also appears in the
OPE of $\phi^{a}\times\bar{\phi}^{b}$.

\subsection{Leading correction to the OPE in the mass expansion}

In this section we compute certain OPE coefficients by considering
the mass expansion of the two-point function. We follow the approach
of section \ref{sec:corrections-to-OPE}, i.e., we shall consider
the two-point function close to criticality and compare its mass expansion
with the OPE. By doing so, we are able to compute the OPE coefficient
associated with operators that have a non-zero VEV.

It is straightforward to check that $\left[\frac{\phi^{a}\phi^{b}}{2}\right]_{T}$
has a vanishing VEV. 
Therefore, we expect that it is not possible
to calculate the associated OPE coefficient by looking at the mass
expansion of the two-point function. Thus, in the following we will
focus mostly on $\left[\frac{\phi^{a}\phi^{b}}{2}\right]_{S}$. 
Since $\left[\frac{\phi^{a}\phi^{b}}{2}\right]_{S}$ is proportional to $\left[\frac{\phi^{m}\phi^{m}}{2}\right]$,
we shall work with the latter from now on.

\subsubsection{The composite operator $\left[\frac{\phi^{m}\phi^{m}}{2}\right]$}

In this section we focus on the VEV and on the
normalization of $\left[\frac{\phi^{m}\phi^{m}}{2}\right]$ and compare
the result against the $N=1$ case of section \ref{subsec:The-composite-operator-phi2}. 

Before considering the VEV of $\left[\frac{\phi^{m}\phi^{m}}{2}\right]$,
let us mention that it is straightforward to compute the anomalous
dimension of $\left[\frac{\phi^{a}\phi^{b}}{2}\right]_{S}$ and $\left[\frac{\phi^{a}\phi^{b}}{2}\right]_{T}$,
which reads $\gamma_{S}=\frac{N+2}{3}\frac{g}{\left(4\pi\right)^{2}}+O\left(\epsilon^{2}\right)$
and $\gamma_{T}=\frac{2}{3}\frac{g}{\left(4\pi\right)^{2}}+O\left(\epsilon^{2}\right)$,
respectively. At the fixed point, the associated scaling dimensions
then read $\Delta_{S}\approx2-\frac{6\epsilon}{N+8}$ and $\Delta_{T}\approx2-\frac{N+6}{N+8}\epsilon$,
respectively. (See \cite{Dey2017a} for a calculation based on the so-called analytic bootstrap.) 

At order $O\left(g^{0}\right)$,
the VEV of $\left[\frac{\phi^{m}\phi^{m}}{2}\right]$
is equivalent to the sum of $N$ VEVs of the single field $\left[\frac{\phi^{2}}{2}\right]$ of the $N=1$ case. 
At order $O\left(g\right)$, a straightforward
calculation shows that all the diagrams and counterterms are the same
as the $N=1$ case but are multiplied by a factor $N\frac{N+2}{3}$.

The amplitude $\langle\left[\frac{\phi^{m}\phi^{m}}{2}\right]\left[\frac{\phi^{n}\phi^{n}}{2}\right]\rangle$
receives fully analogous modifications: the order $O\left(g^{0}\right)$
is equal to the $N$-times the $N=1$ case while the order $O\left(g\right)$
is equal to the $N\frac{N+2}{3}$-times the $N=1$ case.

\subsubsection{OPE coefficients $C_{\phi\phi\left[\frac{\phi^{m}\phi^{m}}{2}\right]}$
and $C_{\phi\bar{\phi}\left[\frac{\phi^{m}\phi^{m}}{2}\right]}$}

To evaluate the coefficients $C_{\phi\phi\left[\frac{\phi^{m}\phi^{m}}{2}\right]}$
and $C_{\phi\bar{\phi}\left[\frac{\phi^{m}\phi^{m}}{2}\right]}$ from
the mass expansion of the propagator we consider the two-point functions
$\langle\phi^{a=1}\phi^{a=1}\rangle$ and $\langle\phi^{a=1}\bar{\phi}^{a=1}\rangle$,
where we fixed the indices to $a=b=1$ without loss of generality.
The two-point function and response function are given by the very
same expression that was considered in section \ref{sec:corrections-to-OPE}
and are given in Appendix \ref{sec:Conventions}. 
The only difference comes from the fact that the mass correction is given by the substitution
(\ref{eq:mass-corr-Og}) with the order $O\left(g\right)$ term being multiplied by $\frac{N+2}{3}$. 
One then finds
\begin{eqnarray}
C_{\phi\phi\left[\frac{\phi^{m}\phi^{m}}{2}\right]} & \approx & \frac{1}{N}\Biggr\{2+g\frac{N+2}{3}\left(\frac{\log\left(\pi x^{2}\mu^{2}\right)+\gamma}{16\pi^{2}}+\frac{1}{\left(4\pi\right)^{2}}\Gamma\left(0,\frac{x^{2}}{4D|t|}\right)\right)\Biggr\}\,,\label{eq:cphiphiphi2_ON-non-normalized}
\end{eqnarray}
where we neglected corrections due to the mass and of higher order
in the interaction coupling.

In full analogy, we can evaluate the coefficient associated with the
response function, it gives
\begin{eqnarray}
C_{\phi\bar{\phi}\left[\frac{\phi^{m}\phi^{m}}{2}\right]} & \approx & -g\frac{N+2}{3N}\frac{\theta\left(t\right)}{\left(4\pi\right)^{d/2}}\frac{1}{\left(Dt\right)}e^{-\frac{x^{2}}{4Dt}}\,.\label{eq:cphiphibphi2_ON-non-normalized}
\end{eqnarray}
Clearly, both expressions (\ref{eq:cphiphiphi2_ON-non-normalized})
and (\ref{eq:cphiphibphi2_ON-non-normalized}) go back to those already
found for $N=1$.

In order to obtain a universal form of these results we must suitably rescale the operators and thereby the coefficients. 
We define normalization
conditions exactly as in section \ref{subsec:Application-to-our-operators}.
By doing so, one obtains
\begin{subequations}
\begin{empheq}[box=\widefbox]{align}
  C_{\phi_{1}\phi_{1}S}\left(y,x\right) & =  \frac{1}{x^{2\Delta_{\phi_{1}}-\Delta_{S}}}\sqrt{\frac{2}{N}}\left(1-\frac{N+2}{2\left(N+8\right)}\epsilon+\frac{N+2}{2\left(N+8\right)}\Gamma\left(0,\frac{1}{|y|}\right)\epsilon\right) \\
C_{\phi_{1}\bar{\phi}_{1}S}\left(y,x\right) & = -\frac{1}{x^{\Delta_{\phi_{1}}+\Delta_{\bar{\phi}_{1}}-\Delta_{S}}}\frac{\left(N+2\right)}{\sqrt{2N}\left(N+8\right)}\frac{e^{1-\frac{1}{y}}\theta\left(y\right)}{y}\epsilon
\end{empheq}
\end{subequations}
up to order $O\left(\epsilon^{2}\right)$.
The static limit of $C_{\phi_{1}\phi_{1}S}$ matches 
with the result found in \cite{Dey2017a} via the analytic conformal bootstrap.

\section{Large N limit\label{sec:Large-N-limit}}

We consider the dynamical $O\left(N\right)$ model in dimension $2<d<4$.

\subsection{Action in the large N limit}

In this section we rewrite the action of the model A in a way that
allows one to take the large $N$ limit straightforwardly. In order
to do so, we translate the SDE from Ito to the Stratonovich interpretation
and consider the associated path integral. 
The reason for this is
that certain formal manipulations are based on the standard rules of calculus, 
which are not always valid for the path integral. 
This is particularly relevant when there are ``ordering ambiguities'', i.e.,
terms such as $\hat{p}\hat{q}$ in quantum mechanics. As we have seen
in section \ref{subsec:Gaussian-Fokker-Planck-Hamiltonian}, the Fokker-Planck
Hamiltonian contains such terms. Ordering ambiguities are in one-to-one
correspondence with the choice of path integral discretization \cite{Berezin:1980xw}
and we shall stick with the midpoint prescription, which is associated
with Weyl ordering. As reviewed in Appendix \ref{sec:FP-Hamiltonian-ordering-midpoint},
by using this discretization an integral over a time derivative satisfies the usual Stokes theorem.

To study the large $N$ limit we follow the approach explained in \cite{Moshe:2003xn}.
In Appendix \ref{app:action_large_N} we show that the large $N$ limit is captured by the following action
\begin{eqnarray}
S & = & N\Biggr\{\int_{tx}\bar{\sigma}\left(\partial_{t}\sigma+D_{0}\left(-\partial^{2}+r_{0}+\frac{u_{0}}{3!}\rho\right)\sigma\right)
\nonumber \\
 &  & -i\int_{tx}\frac{\lambda}{2}\left(\rho-\sigma^{2}\right)-\int_{tx}D_{0}\bar{\sigma}\bar{\sigma}\Biggr]
\nonumber \\
 &  & -\frac{1}{2}D\int_{t,xy}\left(\left(-\partial_{x}^{2}+r\right)+\frac{u_{0}}{3!}\rho\right)\delta\left(x-y\right)
\nonumber \\
 &  & +\frac{1}{2}\mbox{Tr}\log\left(\frac{1}{2D_{0}}\left(\partial_{t}+D_{0}\left(\cdots\right)\right)\left(-\partial_{t}+D_{0}\left(\cdots\right)\right)+i\lambda\right)\Biggr\}\,.
\end{eqnarray}

The functional integral is then evaluated by employing the saddle-point approximation. 
We look for constant solutions to the equation of motion,
which we write below:
\begin{eqnarray}
0 & = & D_{0}\left(r_{0}+\frac{u_{0}}{3!}\rho\right)\sigma-2D_{0}\bar{\sigma}\\
0 & = & D_{0}\left(r_{0}+\frac{u_{0}}{3!}\rho\right)\bar{\sigma}+i\lambda\sigma\\
0 & = & -i\frac{1}{2}\left(\rho-\sigma^{2}\right)+\frac{1}{2}\int_{\omega q}\frac{i}{\frac{\omega^{2}+D_{0}^{2}\left(q^{2}+r_{0}+\frac{u_{0}}{3!}\rho\right)^{2}}{2D_{0}}+i\lambda}\\
0 & = & \bar{\sigma}D_{0}\frac{u_{0}}{3!}\sigma-i\frac{\lambda}{2}-\frac{1}{2}D_{0}\left(\int_{q}\right)\frac{u_{0}}{3!}\\
 &  & +\frac{1}{2}\int_{\omega q}\frac{D_{0}\left(q^{2}+r_{0}+\frac{u_{0}}{3!}\rho\right)}{\left(\frac{\omega^{2}+D_{0}^{2}\left(q^{2}+r_{0}+\frac{u_{0}}{3!}\rho\right)^{2}+2D_{0}i\lambda}{2D_{0}}\right)}\frac{u_{0}}{3!}\,.
\end{eqnarray}
In the present approximation, the constant solutions represent the expectation values of the fluctuating fields themselves. Thus, the solution
for $\bar{\sigma}$ approximate $\langle\bar{\sigma}\rangle$. Since
we know that $\langle\bar{\sigma}\rangle=0$ we look for solutions
such that $\bar{\sigma}=0$. Furthermore, since these constant solutions
represent expectation values of the fields, 
which are time-independent due to time translation invariance, 
we expect that they satisfy the same equation of the static case. 
It is possible to check that for $\lambda=0$
one recovers the static equations. 
In particular, the second and the
fourth equations vanish identically and one is left with
\begin{eqnarray}
0 & = & \left(r_{0}+\frac{u_{0}}{3!}\rho\right)\sigma\\
0 & = & -\frac{1}{2}\left(\rho-\sigma^{2}\right)+\frac{1}{2}\int_{q}\frac{1}{\left(q^{2}+r_{0}+\frac{u_{0}}{3!}\rho\right)}\,.
\end{eqnarray}
It is convenient to define $m^{2}\equiv r_{0}+\frac{u_{0}}{3!}\rho$
so that
\begin{eqnarray}
0 & = & m^{2}\sigma\\
0 & = & \frac{\sigma^{2}}{2}-\frac{3}{u}\left(m^{2}-r_{0}\right)+\frac{1}{2}\int_{q}\frac{1}{\left(q^{2}+m^{2}\right)}\,.
\end{eqnarray}
In the symmetric phase $\sigma=0$ and $m^{2}\neq0$ while for the
broken phase $\sigma\neq0$ and $m^{2}=0$. By employing dim-reg,
the symmetric phase value of $m^{2}$ is determined by
\begin{eqnarray}
0 & = & \frac{3!}{u_{0}}\left(r_{0}-m^{2}\right)+\frac{\Gamma\left(1-d/2\right)}{\left(4\pi\right)^{d/2}}\left(m^{2}\right)^{\frac{d-2}{2}}\,.
\end{eqnarray}
In this case the critical value of the mass is $r_{0,c}=0$ and $m^{2}=0$.
In the symmetric phase and very close to criticality we can approximate
\begin{eqnarray}
m^{2} & \approx & \left(-\frac{\left(4\pi\right)^{d/2}}{\Gamma\left(1-d/2\right)}\frac{3!}{u_{0}}r_{0}\right)^{\frac{2}{d-2}}\,\equiv\,m_{{\rm o.s.}}^{2}\,.
\end{eqnarray}
Note that this allows one to compute straightforwardly the expectation
value $\langle\phi^{2}\rangle=\langle N\rho\rangle\approx N\frac{3!}{u_{0}}\left(m_{{\rm o.s.}}^{2}-r_{0}\right)$.

\subsection{OPE coeffients}

To compute the OPE coefficient, we consider the mass-expansion of the two-point function. 
In order to identify the OPE coefficients,
we assume that in this expansion the non-analytic dependence on $r_{0}$ is due to the expectation value of composite operators.\footnote{The non-analytic dependence on $u_{0}$ is instead expected from the
form that the action can take in the static limit, see \cite{Rose:2021zdk} and
references therein.}

In the large $N$ limit one obtains
\begin{eqnarray}
\langle\sigma\sigma\rangle & \approx & \frac{1}{N}\frac{1}{\left(4\pi\right)^{d/2}}\Biggr\{2^{d-2}\Gamma\left(\frac{d}{2}-1\right)\frac{1}{x^{d-2}}+\Gamma\left(1-\frac{d}{2}\right)m_{{\rm o.s.}}^{d-2}-2^{d-4}\Gamma\left(\frac{d}{2}-2\right)x^{4-d}m_{{\rm o.s.}}^{2}
\nonumber \\
 &  & -\frac{2^{d-2}}{x^{d-2}}\Gamma\left(\frac{d}{2}-1,\frac{x^{2}}{4D|t|}\right)+\frac{2^{d-4}}{x^{d-2}}m_{{\rm o.s.}}^{2}x^{2}\Gamma\left(\frac{d}{2}-2,\frac{x^{2}}{4D|t|}\right)\Biggr\}\,.
\end{eqnarray}
The part non-analytic in $r_{0}$ is due to $m_{{\rm o.s.}}^{2}$,
which is related to $\frac{u_{0}}{3!N}\langle\phi^{2}\rangle$. Thus,
we can read off the OPE coefficient and normalize the operator in
the same way as we did in section \ref{sec:Universal-quantities-from-OPE}.
The normalization of $\phi_{2}$ has already been computed in the
static theory in \cite{Rose:2021zdk},
to which we refer for the detailed calculation
of the normalization. 
To the leading order in $1/N$, one finally obtains
\begin{empheq}[box=\widefbox]{equation}
C_{\phi_{1}\phi_{1}\phi_{2}} = \frac{1}{x^{2\Delta_{1}-\Delta_{2}}}\sqrt{\frac{2}{N}}\sqrt{\frac{\Gamma\left(d-2\right)\sin\left(\pi\frac{d}{2}\right)}{\left(\frac{d}{2}-2\right)\pi}}\frac{1}{\Gamma\left(\frac{d}{2}-1\right)}\left[1-\frac{\Gamma\left(\frac{d}{2}-2,\frac{1}{|y|}\right)}{\Gamma\left(\frac{d}{2}-2\right)}\right]\,.
\end{empheq}

In full analogy, one can determine the OPE coefficient appearing in
the response function. After the normalization of the operators it takes the form
\begin{empheq}[box=\widefbox]{equation}
C_{\phi_{1}\bar{\phi}_{1}\phi_{2}} = -\frac{2}{\sqrt{N}\sqrt{\pi}}\frac{\theta\left(y\right)}{x^{d-2}}e^{1-\frac{1}{y}}y^{1-\frac{d}{2}}\sqrt{\frac{\sin\left(\frac{\pi d}{2}\right)\Gamma\left(d-2\right)}{\left(d-4\right)\Gamma\left(\frac{d}{2}-2\right)^{2}}}
\end{empheq}
to the leading order in $1/N$.

\section{Summary and conclusions \label{sec:Summary-and-conclusions}}

In this work we considered the OPE in the context of critical dynamics
and computed some OPE coefficients for the so-called model A. 
More precisely, we have calculated the leading OPE coefficients of the two-point function and the response function in the case where the fluctuation-dissipation theorem holds.
Moreover,
we showed how it is possible to compute certain universal quantities
that are related to the OPE coefficients by some suitable normalizations.
We calculated explicitly such universal quantities in several cases and approximations. 
These quantities and our discussion of
the role of the fluctuation-dissipation theorem for the calculated
coefficients constitute the main results of the present work. 
Since the OPE in the dynamical case has been far less explored than its static counterpart,
our work constitutes an exploration of the use of the OPE framework in critical dynamics.

More precisely, in section \ref{sec:Gaussian-OPE} we derived the
Gaussian form of the OPE. We found it instructive to derive it by
introducing a normal ordering prescription stemming from interpreting
the Fokker-Planck equation as a functional ``Schroedinger'' equation.
Within the $\epsilon$-expansion,
the leading corrections to the Gaussian OPE have been calculated in section
\ref{sec:corrections-to-OPE} by considering a mass expansion of the
correlation function of interest. By defining suitable normalizations
we have derived a universal form of the OPE coefficients in section
\ref{sec:Universal-quantities-from-OPE}. The $O\left(N\right)$ dynamical
model has been considered in sections \ref{sec:Extention-to-ON} and
\ref{sec:Large-N-limit} where the perturbation theory and the large
$N$ limit have been considered.

It is possible to extend the present study in several directions.
On the one hand, it is possible to push our calculations to higher
order in the coupling expansion
and to consider further OPE coefficients. 
On the other hand,
it would be interesting to extend our analysis
to other dynamical models, 
such as those considered in the Hohenberg-Halperin classification \cite{Hohenberg:1977ym}. 
We do not see any particular obstacle preventing this generalization and we expect that certain aspects, such as the application of FDT, are fully analogous.
A natural questions to be addressed here is if any particular constraint applies to the OPE of conserved modes.
Furthermore,
the present paper considers the critical dynamics associated with systems that have already relaxed to equilibrium. 
Therefore, it would be interesting to generalize our study also to the case of non-thermal
fixed points \cite{Berges:2015kfa} or to consider the effect of the initial
state in more detail \cite{Taeuber_2014}.
It would also be very interesting to understand if there is any method that allows one to determine the OPE coefficients
``self-consistently'', i.e.,
by imposing a number of suitable requirements.
This is, however, beyond the scope of the present paper.
Finally,
the reader may wonder if any of the present results are relevant for experiments.
Let us note that the used approximations, 
i.e., first order in the $\epsilon$-expansion and leading order in the $1/N$-expansion,
are too rough to be of any experimental interest. 
However, if a sufficiently precise result for the coefficients and for the expectation values of the composite operator was obtained, 
one could in principle compare the results with experiments or simulations by looking at the two-point functions slightly away from the critical temperature or at the four-point function in a suitable momentum configuration.

\section*{Acknowledgments}

C.~P.~thanks prof.~Andrea Gambassi for useful discussions.

\newpage{}

\appendix

\section{Conventions \label{sec:Conventions}}

We employ the following short-hand notations. In real space
\begin{eqnarray}
\int_{t}\equiv\int dt & \mbox{ and } & \int_{x}\equiv\int d^{d}x\,,
\end{eqnarray}
while in momentum space
\begin{eqnarray}
\int_{\omega}\equiv\int\frac{d\omega}{2\pi} & \mbox{ and } & \int_{q}\equiv\int\frac{d^{d}q}{\left(2\pi\right)^{d}}\,.
\end{eqnarray}
The Fourier and inverse Fourier transforms are defined as follows:
\begin{eqnarray}
f\left(t,x\right) & = & \int_{\omega q}e^{-i\omega t+iqx}\tilde{f}\left(\omega,q\right)\\
\tilde{f}\left(\omega,q\right) & = & \int_{tx}e^{+i\omega t-iqx}f\left(t,x\right)\,.
\end{eqnarray}

The two-point function in the free-field theory approximation are
given by the following expressions:
\begin{eqnarray}
\langle\phi\left(t,x\right)\phi\left(0,0\right)\rangle & = & \int_{\omega q}e^{-i\omega t+iqx}\frac{2D}{\omega^{2}+D^{2}\left(q^{2}+m^{2}\right)^{2}}\nonumber \\
 & = & \frac{1}{\left(4\pi\right)^{d/2}}2^{d/2}\frac{1}{x^{d-2}}\left(m^{2}x^{2}\right)^{\frac{d-2}{4}}K_{\frac{d-2}{2}}\left(\sqrt{m^{2}x^{2}}\right)\nonumber \\
 &  & \frac{1}{\left(4\pi\right)^{d/2}}\int_{0}^{D}ds\left(-\left|t\right|\frac{1}{\left(4\pi\right)^{d/2}}\left(st\right)^{-d/2}\exp\left[-\frac{x^{2}}{4s\left|t\right|}-s\left|t\right|m^{2}\right]\right)\nonumber \\
& = & \frac{1}{\left(4\pi\right)^{d/2}}\frac{1}{x^{d-2}}\Biggr\{2^{d/2}\left(m^{2}x^{2}\right)^{\frac{d-2}{4}}K_{\frac{d-2}{2}}\left(\sqrt{m^{2}x^{2}}\right)
\label{eq:Gphiphi-Gaussian} \\
 &  & -2^{d-2}\Gamma\left(\frac{d}{2}-1,\frac{x^{2}}{4D|t|}\right)+2^{d-4}m^{2}x^{2}\Gamma\left(\frac{d}{2}-2,\frac{x^{2}}{4D|t|}\right)\Biggr\}+o\left(m^{2}\right),
 \nonumber
\end{eqnarray}
 and
\begin{eqnarray}
\langle\phi\left(t,x\right)\bar{\phi}\left(0,0\right)\rangle & = & \int_{\omega}\int_{q}e^{-i\omega t}e^{iqx}\frac{1}{-i\omega+D\left(q^{2}+m^{2}\right)}\nonumber \\
 & = & \theta\left(t\right)\frac{1}{\left(4\pi\right)^{d/2}}\frac{1}{\left(Dt\right)^{d/2}}e^{-m^{2}Dt-\frac{x^{2}}{4Dt}}\,.\label{eq:eq:Gphiphib-Gaussian}
\end{eqnarray}

\section{Details regarding the composite operator $\left[\phi^2/2\right]$
\label{app:details_phi2}}

In this section we provide details regarding the calculations and definitions associated with $\left[\phi^2/2\right]$.

\begin{figure}
\begin{subfigure}{.3\textwidth}
\centering
\includegraphics[scale=.9]{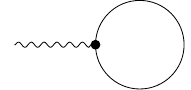}
\caption{$\,$} \label{fig:sfig1}
\end{subfigure} \hfill
\begin{subfigure}{.3\textwidth}
\centering
\includegraphics[width=.8\linewidth]{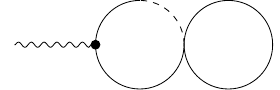}
\caption{$\,$} \label{fig:sfig2}
\end{subfigure} \hfill
\begin{subfigure}{.3\textwidth}
\centering
\includegraphics[width=.8\linewidth]{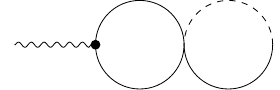}
\caption{$\,$} \label{fig:sfig3}
\end{subfigure}
\caption{The diagram \ref{fig:sfig1} is the diagram contributing to the VEV of $\left[\phi^2/2\right]$ at Gaussian order. The diagrams \ref{fig:sfig2}  contributes at order $O(g)$ while diagram \ref{fig:sfig3} vanishes. We do not display here the diagrams associated with the counterterms stemming out of the renormalization factors $Z_{22}$, $Z_{20}$, and $Z_{m^2}$.} \label{fig:phi2-VEV}
\end{figure}

At the Gaussian level, only the VEV of the operator requires renormalization.
By demanding that $\langle\left[\phi^{2}/2\right]\rangle$ be finite
one obtains
\begin{eqnarray}
Z_{20} & = & \frac{1}{16\pi^{2}\epsilon}\,.
\end{eqnarray}
The diagram associated with $\langle\left[\phi^{2}/2\right]\rangle$
is shown in Figure \ref{fig:phi2-VEV}.

\begin{figure}
\centering
\includegraphics[scale=1]{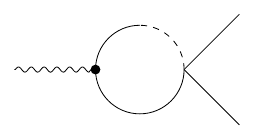}
\caption{The diagram is associated with the three-point function $\langle \phi \phi \left[ \frac{\phi^2}{2}\right]\rangle$ and determines $Z_{22,1}$.} \label{fig:gamma21-Og}
\end{figure}

The mixing matrix $Z_{ij}$ can be expanded in a power series with
respect to the coupling constant $g$. We employ the following notation:
\begin{eqnarray}
Z_{ij} & = & \sum_{k}Z_{ij,k}g^{k}\,.
\end{eqnarray}
At the Gaussian level, one finds $Z_{22,0}=1$ and $Z_{20,0}=\frac{1}{16\pi^{2}\epsilon}$.
To order $O\left(g\right)$ one obtains the following result for the
counterterms
\begin{eqnarray}
Z_{22,1} & = & \frac{1}{\left(4\pi\right)^{2}}\frac{1}{\epsilon}\\
Z_{20,1} & = & \frac{1}{256\pi^{4}\epsilon^{2}}\,.
\end{eqnarray}
The counterterm $Z_{22,1}$ is determined by considering the diagram
in Figure \ref{fig:gamma21-Og}. The counterterm $Z_{22,1}$ implies the anomalous dimension $\gamma_{\phi^{2}}=-Z_{\phi^{2}}^{-1}\dot{Z}_{\phi^{2}}\approx\frac{g}{16\pi^{2}}$.
By using the fixed point value $g_{*}\approx16\pi^{2}\epsilon/3$,
one obtains the scaling dimension $\Delta_2$
given in (\ref{eq:Delta_2_O_epsilon}).

The $O\left(g\right)$-correction to the VEV $\langle\left[\phi^{2}/2\right]\rangle$
is determined by the diagrams shown in Figure \ref{fig:phi2-VEV}
together with those involving the relevant counterterms. 
In the presence of translational invariance, which applies to our setting, we expect
the VEV to be the same as in the static case and not to depend on
the coupling $D$. It can be checked indeed that the VEV is the
same in the static and in the dynamical theory.

\begin{figure}
\begin{subfigure}{.3\textwidth}
\centering
\includegraphics[scale=.9]{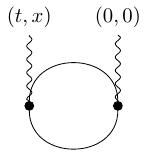}
\caption{$\,$} 
\end{subfigure} \hfill
\begin{subfigure}{.3\textwidth}
\centering
\includegraphics[width=.8\linewidth]{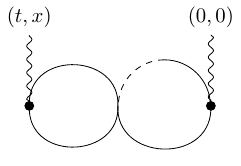}
\caption{$\,$} 
\end{subfigure} \hfill
\begin{subfigure}{.3\textwidth}
\centering
\includegraphics[width=.8\linewidth]{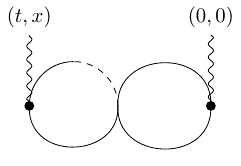}
\caption{$\,$} 
\end{subfigure}
\caption{The diagrams contributing to $\langle \left[\frac{\phi^2}{2}\right] \left(t,x \right) \phi \left[ \frac{\phi^2}{2}\left(0,0 \right)\right] \rangle$ up to  $O(g)$. The diagrams associated with the counterterms are not displayed.} \label{fig:gamma02-Og}
\end{figure}

Finally, we consider the static limit of the two-point of
$\left[\phi^{2}/2\right]$ to $O\left(g\right)$. 
At the fixed point,
the diagrams contributing to this order are shown in Figure \ref{fig:gamma02-Og}.
After an explicit computation and performing the $\epsilon$-expansion
to the first order
at the fixed point, one obtains:
\begin{eqnarray}
{\cal N}_{2} & \approx & 4\sqrt{2}\pi^{2}-\epsilon\frac{2}{3}\sqrt{2}\left(2\gamma\pi^{2}-\pi^{2}+2\pi^{2}\log\pi-2\pi^{2}\log\mu\right)\,.
\end{eqnarray}

\section{Fokker-Planck Hamiltonian, ordering, and mid-point prescription \label{sec:FP-Hamiltonian-ordering-midpoint}}

For the purposes of this appendix, it suffices to consider a generic one-dimensional FP equation:
\begin{eqnarray}
\partial_{t}p\left(x,t\right) & = & -\partial_{x}\left(A\left(x\right)p\left(x,t\right)\right)+D\partial_{x}^{2}p\left(x,t\right)\,.
\end{eqnarray}
By introducing $\hat{p}\equiv-i\partial_{x}$ we can rewrite
\begin{eqnarray}
\partial_{t}p\left(x,t\right) & = & \left[-i\hat{p}A\left(\hat{x}\right)-D\hat{p}^{2}\right]p\left(x,t\right)\\
 & = & -\left[i\hat{p}A\left(\hat{x}\right)+D\hat{p}^{2}\right]p\left(x,t\right)\,.
\end{eqnarray}
As it is written, the ordering is of the $\hat{p}\hat{x}$-type and
if one proceeds to construct the associated path integral one obtains
the ``Ito action'', which is assiociated with the pre-point discretization.
The same FP equation can be rewritten by introducing
\begin{eqnarray}
\left[\hat{p}A\left(\hat{x}\right)\right]_{{\rm Weyl}} & \equiv & \frac{\hat{p}A\left(\hat{x}\right)+A\left(\hat{x}\right)\hat{p}}{2}\,.
\end{eqnarray}
One finds
\begin{eqnarray}
\partial_{t}p\left(x,t\right) & = & -\left[i\left[\hat{p}A\left(\hat{x}\right)\right]_{{\rm Weyl}}+\frac{1}{2}A^{\prime}\left(\hat{x}\right)+D\hat{p}^{2}\right]p\left(x,t\right)\,.
\end{eqnarray}
If one constructs the path integral associated with the above equation
the associated action is composed by a part which is fully similar
to the ``Ito-action'' (but whose associated discretization is dictated
by the midpoint rule) and by a new term, which corresponds to the
second term in the brackets. Such term precisely gives rise to the
well-known determinant. Thus, this clarifies the presence of the determinant
when using the midpoint discretization.

Let us now discuss the role of the midpoint discretization when dealing
with surface terms, see \cite{Chaichian:2001cz}.
In particular, one
wishes to use the standard relation
\begin{eqnarray}
\int_{t_{0}}^{t_{f}}dt\,\frac{d}{dt}f\left(x\left(t\right)\right) & = & f\left(x\left(t_{f}\right)\right)-f\left(x\left(t_{0}\right)\right)\,.
\end{eqnarray}
If one takes the discretized point to correspond to $\left(1-\lambda\right)x_{j}+\lambda x_{j+1}$
with $\lambda\in\left[0,1\right]$ one recovers the prepoint case
for $\lambda=0$ and the midpoint case for $\lambda=1/2$. By denoting
$\delta\equiv x_{j+1}-x_{j}$, the above surface term can be expressed
as follows
\begin{eqnarray}
f\left(x\left(t_{f}\right)\right)-f\left(x\left(t_{0}\right)\right)\,=\,\sum_{j}
\left[ f\left(x_{j+1}\right)-f\left(x_{j}\right) \right] & =\\
\sum_{j}f^{\prime}\left(x_{j}+\lambda\delta\right)\delta+\sum_{j}\frac{1}{2}\left(1-2\lambda\right)f^{\prime\prime}\left(x_{j}\right)\delta^{2}+O\left(\delta^{3}\right)\,.
\end{eqnarray}
For the midpoint choice, i.e., $\lambda=1/2$ the second term does
not appear. However, such terms are generally relevant in the path
integral. Indeed we know that the path integral is determined by the
paths which are continuous but nowhere differentiable implying that
$\delta^2 \sim\Delta t$ and so that the second term gives rise to a
finite term in the action in the continuum limit. To avoid dealing
with such terms we adopt the midpoint discretization.

\section{Action in the large $N$ limit \label{app:action_large_N}}

In this section we rewrite the functional integral associated with the dynamical $O(N)$ model in a way that allows to take the large $N$ limit straightforwardly,
see \cite{Moshe:2003xn}.
The partition function takes the form
\begin{eqnarray}
Z & = & \int{\cal D}\phi{\cal D}\bar{\phi}\exp\Biggr[-\int_{tx}\bar{\phi}^{a}\left(\partial_{t}\phi^{a}+D_{0}\left(-\partial^{2}+r_{0}+\frac{u_{0}}{3!N}\phi^{m}\phi^{m}\right)\phi^{a}\right)+\int_{tx}D_{0}\bar{\phi}^{a}\bar{\phi}^{a}\Biggr]
\nonumber \\
 &  & \times\exp\Biggr[\frac{1}{2}D\frac{\delta^{2}H}{\delta\phi^{a}\delta\phi^{a}}\Biggr]\,
\end{eqnarray}
where the last term arises from the non-trivial functional determinant
present when using the midpoint discretization (see, e.g., \cite{Taeuber_2014}). It reads
\begin{eqnarray}
\exp\Biggr[\frac{1}{2}D\frac{\delta^{2}H}{\delta\phi^{a}\delta\phi^{a}}\Biggr] & = & \exp\Biggr[\frac{1}{2}D\int_{t,xy}\left(N\left(-\partial_{x}^{2}+r\right)+\frac{u_{0}}{3!}\left(\phi^{2}+\frac{2}{N}\phi^{2}\right)\right)\delta\left(x-y\right)\Biggr]\,.
\end{eqnarray}
We rewrite the partition function by inserting a unit factor via a
delta function:
\begin{eqnarray}
Z & = & \int{\cal D}\phi{\cal D}\bar{\phi}\exp\Biggr[-\int_{tx}\bar{\phi}^{a}\left(\partial_{t}\phi^{a}+D_{0}\left(-\partial^{2}+r_{0}+\frac{u_{0}}{3!N}\phi^{m}\phi^{m}\right)\phi^{a}\right)+\int_{tx}D_{0}\bar{\phi}^{a}\bar{\phi}^{a}\Biggr]
\nonumber \\
 &  & \int{\cal D}\lambda{\cal D}\rho\,e^{i\lambda\left(\rho-\phi^{2}\right)}\exp\Biggr[\frac{1}{2}D\frac{\delta^{2}H}{\delta\phi^{a}\delta\phi^{a}}\Biggr]
\nonumber \\
 & = & \int{\cal D}\phi{\cal D}\bar{\phi}\exp\Biggr[-\int_{tx}\bar{\phi}^{a}\left(\partial_{t}\phi^{a}+D_{0}\left(-\partial^{2}+r_{0}+\frac{u_{0}}{3!N}\rho\right)\phi^{a}\right)+\int_{tx}D_{0}\bar{\phi}^{a}\bar{\phi}^{a}\Biggr]
\nonumber \\
 &  & \int{\cal D}\lambda{\cal D}\rho\,e^{i\lambda\left(\rho-\phi^{2}\right)}\exp\Biggr[\frac{1}{2}D\frac{\delta^{2}H}{\delta\phi^{a}\delta\phi^{a}}\left(\rho\right)\Biggr]\,.
\end{eqnarray}
Next, we split the fields: $\phi^{a}=\left(\sigma,\pi^{i}\right)$
and $\bar{\phi}^{a}=\left(\bar{\sigma},\bar{\pi}^{i}\right)$ and
integrate over $\pi$ and $\bar{\pi}$. One obtains the following
action
\begin{eqnarray}
S & = & \int_{tx}\bar{\sigma}\left(\partial_{t}\sigma+D_{0}\left(-\partial^{2}+r_{0}+\frac{u_{0}}{3!N}\rho\right)\sigma\right)
\nonumber \\
 &  & -i\int_{tx}\lambda\left(\rho-\sigma^{2}\right)-\int_{tx}D_{0}\bar{\sigma}\bar{\sigma}\Biggr]-\frac{1}{2}D\frac{\delta^{2}H}{\delta\phi^{a}\delta\phi^{a}}\left(\rho\right)
\nonumber \\
 &  & +\frac{N-1}{2}\mbox{Tr}\log\left(\frac{1}{2D_{0}}\left(\partial_{t}+D_{0}\left(\cdots\right)\right)\left(-\partial_{t}+D_{0}\left(\cdots\right)\right)+2i\lambda\right)\,.
\end{eqnarray}
To make the large $N$ limit manifest we make some rescalings: $\left(\sigma,\bar{\sigma}\right)\rightarrow\sqrt{N}\left(\sigma,\bar{\sigma}\right)$,
$\lambda\rightarrow\frac{\lambda}{2}$, and $\rho\rightarrow N\rho$
so that the action now reads
\begin{eqnarray}
S & = & N\Biggr\{\int_{tx}\bar{\sigma}\left(\partial_{t}\sigma+D_{0}\left(-\partial^{2}+r_{0}+\frac{u_{0}}{3!}\rho\right)\sigma\right)
\nonumber \\
 &  & -i\int_{tx}\frac{\lambda}{2}\left(\rho-\sigma^{2}\right)-\int_{tx}D_{0}\bar{\sigma}\bar{\sigma}\Biggr]
\nonumber \\
 &  & -\frac{1}{2}D\int_{t,xy}\left(\left(-\partial_{x}^{2}+r\right)+\frac{u_{0}}{3!}\rho\right)\delta\left(x-y\right)
\nonumber \\
 &  & +\frac{1}{2}\mbox{Tr}\log\left(\frac{1}{2D_{0}}\left(\partial_{t}+D_{0}\left(\cdots\right)\right)\left(-\partial_{t}+D_{0}\left(\cdots\right)\right)+i\lambda\right)\Biggr\}\,,
\end{eqnarray}
where we also used $N-1\approx N$ and we neglected subleading terms
in the $1/N$ expansion of the functional determinant.

\bibliography{paper}

\begin{thebibliography}{45}%
\makeatletter
\providecommand \@ifxundefined [1]{%
 \@ifx{#1\undefined}
}%
\providecommand \@ifnum [1]{%
 \ifnum #1\expandafter \@firstoftwo
 \else \expandafter \@secondoftwo
 \fi
}%
\providecommand \@ifx [1]{%
 \ifx #1\expandafter \@firstoftwo
 \else \expandafter \@secondoftwo
 \fi
}%
\providecommand \natexlab [1]{#1}%
\providecommand \enquote  [1]{``#1''}%
\providecommand \bibnamefont  [1]{#1}%
\providecommand \bibfnamefont [1]{#1}%
\providecommand \citenamefont [1]{#1}%
\providecommand \href@noop [0]{\@secondoftwo}%
\providecommand \href [0]{\begingroup \@sanitize@url \@href}%
\providecommand \@href[1]{\@@startlink{#1}\@@href}%
\providecommand \@@href[1]{\endgroup#1\@@endlink}%
\providecommand \@sanitize@url [0]{\catcode `\\12\catcode `\$12\catcode
  `\&12\catcode `\#12\catcode `\^12\catcode `\_12\catcode `\%12\relax}%
\providecommand \@@startlink[1]{}%
\providecommand \@@endlink[0]{}%
\providecommand \url  [0]{\begingroup\@sanitize@url \@url }%
\providecommand \@url [1]{\endgroup\@href {#1}{\urlprefix }}%
\providecommand \urlprefix  [0]{URL }%
\providecommand \Eprint [0]{\href }%
\providecommand \doibase [0]{http://dx.doi.org/}%
\providecommand \selectlanguage [0]{\@gobble}%
\providecommand \bibinfo  [0]{\@secondoftwo}%
\providecommand \bibfield  [0]{\@secondoftwo}%
\providecommand \translation [1]{[#1]}%
\providecommand \BibitemOpen [0]{}%
\providecommand \bibitemStop [0]{}%
\providecommand \bibitemNoStop [0]{.\EOS\space}%
\providecommand \EOS [0]{\spacefactor3000\relax}%
\providecommand \BibitemShut  [1]{\csname bibitem#1\endcsname}%
\let\auto@bib@innerbib\@empty
\bibitem [{\citenamefont {Wilson}(1969)}]{Wilson:1969zs}%
  \BibitemOpen
  \bibfield  {author} {\bibinfo {author} {\bibfnamefont {K.~G.}\ \bibnamefont
  {Wilson}},\ }\bibfield  {title} {\enquote {\bibinfo {title} {{Nonlagrangian
  models of current algebra}},}\ }\href {\doibase 10.1103/PhysRev.179.1499}
  {\bibfield  {journal} {\bibinfo  {journal} {Phys. Rev.}\ }\textbf {\bibinfo
  {volume} {179}},\ \bibinfo {pages} {1499--1512} (\bibinfo {year}
  {1969})}\BibitemShut {NoStop}%
\bibitem [{\citenamefont {Zimmermann}(1973)}]{Zimmermann:1972tv}%
  \BibitemOpen
  \bibfield  {author} {\bibinfo {author} {\bibfnamefont {Wolfhart}\
  \bibnamefont {Zimmermann}},\ }\bibfield  {title} {\enquote {\bibinfo {title}
  {{Normal products and the short distance expansion in the perturbation theory
  of renormalizable interactions}},}\ }\href {\doibase
  10.1016/0003-4916(73)90430-2} {\bibfield  {journal} {\bibinfo  {journal}
  {Ann. Phys. (N. Y.)}\ }\textbf {\bibinfo {volume} {77}},\ \bibinfo {pages}
  {570--601} (\bibinfo {year} {1973})}\BibitemShut {NoStop}%
\bibitem [{\citenamefont {Francesco}\ \emph {et~al.}(1997)\citenamefont
  {Francesco}, \citenamefont {Mathieu},\ and\ \citenamefont
  {S{\'{e}}n{\'{e}}chal}}]{Francesco1997a}%
  \BibitemOpen
  \bibfield  {author} {\bibinfo {author} {\bibfnamefont {Philippe~Di}\
  \bibnamefont {Francesco}}, \bibinfo {author} {\bibfnamefont {Pierre}\
  \bibnamefont {Mathieu}}, \ and\ \bibinfo {author} {\bibfnamefont {David}\
  \bibnamefont {S{\'{e}}n{\'{e}}chal}},\ }\href {\doibase
  10.1007/978-1-4612-2256-9} {\emph {\bibinfo {title} {Conformal Field
  Theory}}}\ (\bibinfo  {publisher} {Springer New York},\ \bibinfo {year}
  {1997})\BibitemShut {NoStop}%
\bibitem [{\citenamefont {{Poland}}\ \emph {et~al.}(2019)\citenamefont
  {{Poland}}, \citenamefont {{Rychkov}},\ and\ \citenamefont
  {{Vichi}}}]{Poland2019a}%
  \BibitemOpen
  \bibfield  {author} {\bibinfo {author} {\bibfnamefont {David}\ \bibnamefont
  {{Poland}}}, \bibinfo {author} {\bibfnamefont {Slava}\ \bibnamefont
  {{Rychkov}}}, \ and\ \bibinfo {author} {\bibfnamefont {Alessandro}\
  \bibnamefont {{Vichi}}},\ }\bibfield  {title} {\enquote {\bibinfo {title}
  {{The conformal bootstrap: Theory, numerical techniques, and
  applications}},}\ }\href {\doibase 10.1103/RevModPhys.91.015002} {\bibfield
  {journal} {\bibinfo  {journal} {Rev. Mod. Phys.}\ }\textbf {\bibinfo {volume}
  {91}},\ \bibinfo {eid} {015002} (\bibinfo {year} {2019})},\ \Eprint
  {http://arxiv.org/abs/1805.04405} {arXiv:1805.04405 [hep-th]} \BibitemShut
  {NoStop}%
\bibitem [{\citenamefont {Brezin}\ \emph {et~al.}(1974)\citenamefont {Brezin},
  \citenamefont {Amit},\ and\ \citenamefont {Zinn-Justin}}]{Brezin:1974zz}%
  \BibitemOpen
  \bibfield  {author} {\bibinfo {author} {\bibfnamefont {E.}~\bibnamefont
  {Brezin}}, \bibinfo {author} {\bibfnamefont {D.~J.}\ \bibnamefont {Amit}}, \
  and\ \bibinfo {author} {\bibfnamefont {J.}~\bibnamefont {Zinn-Justin}},\
  }\bibfield  {title} {\enquote {\bibinfo {title} {{Next-to-Leading Terms in
  the Correlation Function inside the Scaling Regime}},}\ }\href {\doibase
  10.1103/PhysRevLett.32.151} {\bibfield  {journal} {\bibinfo  {journal} {Phys.
  Rev. Lett.}\ }\textbf {\bibinfo {volume} {32}},\ \bibinfo {pages} {151--154}
  (\bibinfo {year} {1974})}\BibitemShut {NoStop}%
\bibitem [{\citenamefont {Br\'ezin}\ \emph {et~al.}(1974)\citenamefont
  {Br\'ezin}, \citenamefont {le~Guillou},\ and\ \citenamefont
  {Zinn-Justin}}]{Brezin1974b}%
  \BibitemOpen
  \bibfield  {author} {\bibinfo {author} {\bibfnamefont {E.}~\bibnamefont
  {Br\'ezin}}, \bibinfo {author} {\bibfnamefont {J.~C.}\ \bibnamefont
  {le~Guillou}}, \ and\ \bibinfo {author} {\bibfnamefont {J.}~\bibnamefont
  {Zinn-Justin}},\ }\bibfield  {title} {\enquote {\bibinfo {title} {Asymptotic
  behavior of the spin-spin correlation function in a field and below
  ${T}_{c}$},}\ }\href {\doibase 10.1103/PhysRevLett.32.473} {\bibfield
  {journal} {\bibinfo  {journal} {Phys. Rev. Lett.}\ }\textbf {\bibinfo
  {volume} {32}},\ \bibinfo {pages} {473--475} (\bibinfo {year}
  {1974})}\BibitemShut {NoStop}%
\bibitem [{\citenamefont {Caselle}\ \emph {et~al.}(2015)\citenamefont
  {Caselle}, \citenamefont {Costagliola},\ and\ \citenamefont
  {Magnoli}}]{Caselle:2015csa}%
  \BibitemOpen
  \bibfield  {author} {\bibinfo {author} {\bibfnamefont {M.}~\bibnamefont
  {Caselle}}, \bibinfo {author} {\bibfnamefont {G.}~\bibnamefont
  {Costagliola}}, \ and\ \bibinfo {author} {\bibfnamefont {N.}~\bibnamefont
  {Magnoli}},\ }\bibfield  {title} {\enquote {\bibinfo {title} {{Numerical
  determination of the operator-product-expansion coefficients in the 3D Ising
  model from off-critical correlators}},}\ }\href {\doibase
  10.1103/PhysRevD.91.061901} {\bibfield  {journal} {\bibinfo  {journal} {Phys.
  Rev. D}\ }\textbf {\bibinfo {volume} {91}},\ \bibinfo {pages} {061901}
  (\bibinfo {year} {2015})},\ \Eprint {http://arxiv.org/abs/1501.04065}
  {arXiv:1501.04065 [hep-th]} \BibitemShut {NoStop}%
\bibitem [{\citenamefont {Costagliola}(2016)}]{Costagliola:2015ier}%
  \BibitemOpen
  \bibfield  {author} {\bibinfo {author} {\bibfnamefont {Gianluca}\
  \bibnamefont {Costagliola}},\ }\bibfield  {title} {\enquote {\bibinfo {title}
  {{Operator product expansion coefficients of the 3D Ising model with a
  trapping potential}},}\ }\href {\doibase 10.1103/PhysRevD.93.066008}
  {\bibfield  {journal} {\bibinfo  {journal} {Phys. Rev. D}\ }\textbf {\bibinfo
  {volume} {93}},\ \bibinfo {pages} {066008} (\bibinfo {year} {2016})},\
  \Eprint {http://arxiv.org/abs/1511.02921} {arXiv:1511.02921 [hep-th]}
  \BibitemShut {NoStop}%
\bibitem [{\citenamefont {Caselle}\ \emph {et~al.}(2016)\citenamefont
  {Caselle}, \citenamefont {Costagliola},\ and\ \citenamefont
  {Magnoli}}]{Caselle:2016mww}%
  \BibitemOpen
  \bibfield  {author} {\bibinfo {author} {\bibfnamefont {Michele}\ \bibnamefont
  {Caselle}}, \bibinfo {author} {\bibfnamefont {Gianluca}\ \bibnamefont
  {Costagliola}}, \ and\ \bibinfo {author} {\bibfnamefont {Nicodemo}\
  \bibnamefont {Magnoli}},\ }\bibfield  {title} {\enquote {\bibinfo {title}
  {{Conformal perturbation of off-critical correlators in the 3D Ising
  universality class}},}\ }\href {\doibase 10.1103/PhysRevD.94.026005}
  {\bibfield  {journal} {\bibinfo  {journal} {Phys. Rev. D}\ }\textbf {\bibinfo
  {volume} {94}},\ \bibinfo {pages} {026005} (\bibinfo {year} {2016})},\
  \Eprint {http://arxiv.org/abs/1605.05133} {arXiv:1605.05133 [hep-th]}
  \BibitemShut {NoStop}%
\bibitem [{\citenamefont {Novikov}\ \emph {et~al.}(1978)\citenamefont
  {Novikov}, \citenamefont {Okun}, \citenamefont {Shifman}, \citenamefont
  {Vainshtein}, \citenamefont {Voloshin},\ and\ \citenamefont
  {Zakharov}}]{Novikov:1977dq}%
  \BibitemOpen
  \bibfield  {author} {\bibinfo {author} {\bibfnamefont {V.~A.}\ \bibnamefont
  {Novikov}}, \bibinfo {author} {\bibfnamefont {L.~B.}\ \bibnamefont {Okun}},
  \bibinfo {author} {\bibfnamefont {Mikhail~A.}\ \bibnamefont {Shifman}},
  \bibinfo {author} {\bibfnamefont {A.~I.}\ \bibnamefont {Vainshtein}},
  \bibinfo {author} {\bibfnamefont {M.~B.}\ \bibnamefont {Voloshin}}, \ and\
  \bibinfo {author} {\bibfnamefont {Valentin~I.}\ \bibnamefont {Zakharov}},\
  }\bibfield  {title} {\enquote {\bibinfo {title} {{Charmonium and Gluons:
  Basic Experimental Facts and Theoretical Introduction}},}\ }\href {\doibase
  10.1016/0370-1573(78)90120-5} {\bibfield  {journal} {\bibinfo  {journal}
  {Phys. Rept.}\ }\textbf {\bibinfo {volume} {41}},\ \bibinfo {pages} {1--133}
  (\bibinfo {year} {1978})}\BibitemShut {NoStop}%
\bibitem [{\citenamefont {Martin}\ \emph {et~al.}(1973)\citenamefont {Martin},
  \citenamefont {Siggia},\ and\ \citenamefont {Rose}}]{Martin:1973zz}%
  \BibitemOpen
  \bibfield  {author} {\bibinfo {author} {\bibfnamefont {P.~C.}\ \bibnamefont
  {Martin}}, \bibinfo {author} {\bibfnamefont {E.~D.}\ \bibnamefont {Siggia}},
  \ and\ \bibinfo {author} {\bibfnamefont {H.~A.}\ \bibnamefont {Rose}},\
  }\bibfield  {title} {\enquote {\bibinfo {title} {{Statistical Dynamics of
  Classical Systems}},}\ }\href {\doibase 10.1103/PhysRevA.8.423} {\bibfield
  {journal} {\bibinfo  {journal} {Phys. Rev. A}\ }\textbf {\bibinfo {volume}
  {8}},\ \bibinfo {pages} {423--437} (\bibinfo {year} {1973})}\BibitemShut
  {NoStop}%
\bibitem [{\citenamefont {Janssen}(1976)}]{Janssen:1976a}%
  \BibitemOpen
  \bibfield  {author} {\bibinfo {author} {\bibfnamefont {H.~K.}\ \bibnamefont
  {Janssen}},\ }\href@noop {} {\bibfield  {journal} {\bibinfo  {journal} {Z.
  Phys. B}\ }\textbf {\bibinfo {volume} {23}},\ \bibinfo {pages} {377}
  (\bibinfo {year} {1976})}\BibitemShut {NoStop}%
\bibitem [{\citenamefont {de~Dominicis}(1976)}]{deDominicis:1976a}%
  \BibitemOpen
  \bibfield  {author} {\bibinfo {author} {\bibfnamefont {C.}~\bibnamefont
  {de~Dominicis}},\ }\href@noop {} {\bibfield  {journal} {\bibinfo  {journal}
  {J. Phys. Colloques}\ }\textbf {\bibinfo {volume} {37}},\ \bibinfo {pages}
  {247} (\bibinfo {year} {1976})}\BibitemShut {NoStop}%
\bibitem [{\citenamefont {T{\"a}uber}(2014)}]{Taeuber_2014}%
  \BibitemOpen
  \bibfield  {author} {\bibinfo {author} {\bibfnamefont {Uwe~C.}\ \bibnamefont
  {T{\"a}uber}},\ }\href@noop {} {\emph {\bibinfo {title} {Critical Dynamics: A
  Field Theory Approach to Equilibrium and Non-Equilibrium Scaling Behavior}}}\
  (\bibinfo  {publisher} {Cambridge University Press},\ \bibinfo {year}
  {2014})\BibitemShut {NoStop}%
\bibitem [{\citenamefont {Golkar}\ and\ \citenamefont
  {Son}(2014)}]{Golkar:2014mwa}%
  \BibitemOpen
  \bibfield  {author} {\bibinfo {author} {\bibfnamefont {Siavash}\ \bibnamefont
  {Golkar}}\ and\ \bibinfo {author} {\bibfnamefont {Dam~T.}\ \bibnamefont
  {Son}},\ }\bibfield  {title} {\enquote {\bibinfo {title} {{Operator Product
  Expansion and Conservation Laws in Non-Relativistic Conformal Field
  Theories}},}\ }\href {\doibase 10.1007/JHEP12(2014)063} {\bibfield  {journal}
  {\bibinfo  {journal} {JHEP}\ }\textbf {\bibinfo {volume} {12}},\ \bibinfo
  {pages} {063} (\bibinfo {year} {2014})},\ \Eprint
  {http://arxiv.org/abs/1408.3629} {arXiv:1408.3629 [hep-th]} \BibitemShut
  {NoStop}%
\bibitem [{\citenamefont {Shimada}\ and\ \citenamefont
  {Shimada}(2021)}]{Shimada:2021xsv}%
  \BibitemOpen
  \bibfield  {author} {\bibinfo {author} {\bibfnamefont {Hidehiko}\
  \bibnamefont {Shimada}}\ and\ \bibinfo {author} {\bibfnamefont {Hirohiko}\
  \bibnamefont {Shimada}},\ }\bibfield  {title} {\enquote {\bibinfo {title}
  {{Exact four-point function and OPE for an interacting quantum field theory
  with space/time anisotropic scale invariance}},}\ }\href {\doibase
  10.1007/JHEP10(2021)030} {\bibfield  {journal} {\bibinfo  {journal} {JHEP}\
  }\textbf {\bibinfo {volume} {10}},\ \bibinfo {pages} {030} (\bibinfo {year}
  {2021})},\ \Eprint {http://arxiv.org/abs/2107.07770} {arXiv:2107.07770
  [hep-th]} \BibitemShut {NoStop}%
\bibitem [{\citenamefont {Zinn-Justin}(2002)}]{Zinn_book}%
  \BibitemOpen
  \bibfield  {author} {\bibinfo {author} {\bibfnamefont {J.}~\bibnamefont
  {Zinn-Justin}},\ }\href@noop {} {\emph {\bibinfo {title} {{Quantum Field
  Theory and Critical Phenomena}}}}\ (\bibinfo  {publisher} {Fourth Edition,
  Clarendon Press, Oxford},\ \bibinfo {year} {2002})\BibitemShut {NoStop}%
\bibitem [{\citenamefont {Canet}\ and\ \citenamefont
  {Chate}(2007)}]{Canet:2006xu}%
  \BibitemOpen
  \bibfield  {author} {\bibinfo {author} {\bibfnamefont {Leonie}\ \bibnamefont
  {Canet}}\ and\ \bibinfo {author} {\bibfnamefont {Hugues}\ \bibnamefont
  {Chate}},\ }\bibfield  {title} {\enquote {\bibinfo {title} {{Non-perturbative
  Approach to Critical Dynamics}},}\ }\href {\doibase
  10.1088/1751-8113/40/9/002} {\bibfield  {journal} {\bibinfo  {journal} {J.
  Phys.}\ }\textbf {\bibinfo {volume} {40}},\ \bibinfo {pages} {1937--1950}
  (\bibinfo {year} {2007})},\ \Eprint {http://arxiv.org/abs/cond-mat/0610468}
  {arXiv:cond-mat/0610468} \BibitemShut {NoStop}%
\bibitem [{\citenamefont {Sieberer}\ \emph {et~al.}(2015)\citenamefont
  {Sieberer}, \citenamefont {Chiocchetta}, \citenamefont {Gambassi},
  \citenamefont {T\"auber},\ and\ \citenamefont {Diehl}}]{Sieberer:2015hba}%
  \BibitemOpen
  \bibfield  {author} {\bibinfo {author} {\bibfnamefont {L.~M.}\ \bibnamefont
  {Sieberer}}, \bibinfo {author} {\bibfnamefont {A.}~\bibnamefont
  {Chiocchetta}}, \bibinfo {author} {\bibfnamefont {A.}~\bibnamefont
  {Gambassi}}, \bibinfo {author} {\bibfnamefont {U.~C.}\ \bibnamefont
  {T\"auber}}, \ and\ \bibinfo {author} {\bibfnamefont {S.}~\bibnamefont
  {Diehl}},\ }\bibfield  {title} {\enquote {\bibinfo {title} {{Thermodynamic
  Equilibrium as a Symmetry of the Schwinger-Keldysh Action}},}\ }\href
  {\doibase 10.1103/PhysRevB.92.134307} {\bibfield  {journal} {\bibinfo
  {journal} {Phys. Rev. B}\ }\textbf {\bibinfo {volume} {92}},\ \bibinfo
  {pages} {134307} (\bibinfo {year} {2015})},\ \Eprint
  {http://arxiv.org/abs/1505.00912} {arXiv:1505.00912 [cond-mat.stat-mech]}
  \BibitemShut {NoStop}%
\bibitem [{\citenamefont {Onsager}\ and\ \citenamefont
  {Machlup}(1953)}]{OnsagerMachlup1953}%
  \BibitemOpen
  \bibfield  {author} {\bibinfo {author} {\bibfnamefont {L.}~\bibnamefont
  {Onsager}}\ and\ \bibinfo {author} {\bibfnamefont {S.}~\bibnamefont
  {Machlup}},\ }\bibfield  {title} {\enquote {\bibinfo {title} {Fluctuations
  and irreversible processes},}\ }\href {\doibase 10.1103/PhysRev.91.1505}
  {\bibfield  {journal} {\bibinfo  {journal} {Phys. Rev.}\ }\textbf {\bibinfo
  {volume} {91}},\ \bibinfo {pages} {1505--1512} (\bibinfo {year}
  {1953})}\BibitemShut {NoStop}%
\bibitem [{\citenamefont {Calabrese}\ and\ \citenamefont
  {Gambassi}(2004)}]{Calabrese:2004kn}%
  \BibitemOpen
  \bibfield  {author} {\bibinfo {author} {\bibfnamefont {Pasquale}\
  \bibnamefont {Calabrese}}\ and\ \bibinfo {author} {\bibfnamefont {Andrea}\
  \bibnamefont {Gambassi}},\ }\bibfield  {title} {\enquote {\bibinfo {title}
  {{On the definition of a unique effective temperature for non-equilibrium
  critical systems}},}\ }\href {\doibase 10.1088/1742-5468/2004/07/P07013}
  {\bibfield  {journal} {\bibinfo  {journal} {J. Stat. Mech.}\ }\textbf
  {\bibinfo {volume} {0407}},\ \bibinfo {pages} {P07013} (\bibinfo {year}
  {2004})},\ \Eprint {http://arxiv.org/abs/cond-mat/0406289}
  {arXiv:cond-mat/0406289} \BibitemShut {NoStop}%
\bibitem [{\citenamefont {Bonneau}(2009)}]{Bonneau:2009}%
  \BibitemOpen
  \bibfield  {author} {\bibinfo {author} {\bibfnamefont {G.}~\bibnamefont
  {Bonneau}},\ }\bibfield  {title} {\enquote {\bibinfo {title} {{O}perator
  product expansion},}\ }\href {\doibase 10.4249/scholarpedia.8506} {\bibfield
  {journal} {\bibinfo  {journal} {Scholarpedia}\ }\textbf {\bibinfo {volume}
  {4}},\ \bibinfo {pages} {8506} (\bibinfo {year} {2009})},\ \bibinfo {note}
  {revision \#91610}\BibitemShut {NoStop}%
\bibitem [{\citenamefont {Hatfield}(1992)}]{Hatfield:1992rz}%
  \BibitemOpen
  \bibfield  {author} {\bibinfo {author} {\bibfnamefont {B.}~\bibnamefont
  {Hatfield}},\ }\href@noop {} {\emph {\bibinfo {title} {{Quantum field theory
  of point particles and strings}}}}\ (\bibinfo  {publisher} {Perseus Books},\
  \bibinfo {year} {1992})\BibitemShut {NoStop}%
\bibitem [{\citenamefont {Sakita}(1983)}]{Sakita:1983eq}%
  \BibitemOpen
  \bibfield  {author} {\bibinfo {author} {\bibfnamefont {B.}~\bibnamefont
  {Sakita}},\ }\bibfield  {title} {\enquote {\bibinfo {title} {{Stochastic
  quantization}},}\ }in\ \href@noop {} {\emph {\bibinfo {booktitle} {{22nd
  International Conference on High-Energy Physics}}}}\ (\bibinfo {year}
  {1983})\ p.\ \bibinfo {pages} {I.107}\BibitemShut {NoStop}%
\bibitem [{\citenamefont {Peskin}\ and\ \citenamefont
  {Schroeder}(1995)}]{Peskin:1995ev}%
  \BibitemOpen
  \bibfield  {author} {\bibinfo {author} {\bibfnamefont {Michael~E.}\
  \bibnamefont {Peskin}}\ and\ \bibinfo {author} {\bibfnamefont {Daniel~V.}\
  \bibnamefont {Schroeder}},\ }\href@noop {} {\emph {\bibinfo {title} {{An
  Introduction to quantum field theory}}}}\ (\bibinfo  {publisher}
  {Addison-Wesley},\ \bibinfo {address} {Reading, USA},\ \bibinfo {year}
  {1995})\BibitemShut {NoStop}%
\bibitem [{\citenamefont {Collins}(2023)}]{Collins:1984xc}%
  \BibitemOpen
  \bibfield  {author} {\bibinfo {author} {\bibfnamefont {John~C.}\ \bibnamefont
  {Collins}},\ }\href {\doibase 10.1017/9781009401807} {\emph {\bibinfo {title}
  {{Renormalization}}}},\ \bibinfo {series} {Cambridge Monographs on
  Mathematical Physics}, Vol.~\bibinfo {volume} {26}\ (\bibinfo  {publisher}
  {Cambridge University Press},\ \bibinfo {address} {Cambridge},\ \bibinfo
  {year} {2023})\BibitemShut {NoStop}%
\bibitem [{\citenamefont {Tkachov}(1983)}]{Tkachov:1983st}%
  \BibitemOpen
  \bibfield  {author} {\bibinfo {author} {\bibfnamefont {F.~V.}\ \bibnamefont
  {Tkachov}},\ }\bibfield  {title} {\enquote {\bibinfo {title} {{ON THE
  OPERATOR PRODUCT EXPANSION IN THE MS SCHEME}},}\ }\href {\doibase
  10.1016/0370-2693(83)91438-7} {\bibfield  {journal} {\bibinfo  {journal}
  {Phys. Lett. B}\ }\textbf {\bibinfo {volume} {124}},\ \bibinfo {pages}
  {212--216} (\bibinfo {year} {1983})}\BibitemShut {NoStop}%
\bibitem [{\citenamefont {Guida}\ and\ \citenamefont
  {Magnoli}(1996)}]{Guida:1995kc}%
  \BibitemOpen
  \bibfield  {author} {\bibinfo {author} {\bibfnamefont {Riccardo}\
  \bibnamefont {Guida}}\ and\ \bibinfo {author} {\bibfnamefont {Nicodemo}\
  \bibnamefont {Magnoli}},\ }\bibfield  {title} {\enquote {\bibinfo {title}
  {{All order IR finite expansion for short distance behavior of massless
  theories perturbed by a relevant operator}},}\ }\href {\doibase
  10.1016/0550-3213(96)00175-7} {\bibfield  {journal} {\bibinfo  {journal}
  {Nucl. Phys. B}\ }\textbf {\bibinfo {volume} {471}},\ \bibinfo {pages}
  {361--388} (\bibinfo {year} {1996})},\ \Eprint
  {http://arxiv.org/abs/hep-th/9511209} {arXiv:hep-th/9511209} \BibitemShut
  {NoStop}%
\bibitem [{\citenamefont {Amit}(1984)}]{Amit:1984ms}%
  \BibitemOpen
  \bibfield  {author} {\bibinfo {author} {\bibfnamefont {D.~J.}\ \bibnamefont
  {Amit}},\ }\href@noop {} {\emph {\bibinfo {title} {{FIELD THEORY, THE
  RENORMALIZATION GROUP, AND CRITICAL PHENOMENA}}}}\ (\bibinfo  {publisher}
  {World Scientific},\ \bibinfo {year} {1984})\BibitemShut {NoStop}%
\bibitem [{\citenamefont {Brezin}\ \emph {et~al.}(1972)\citenamefont {Brezin},
  \citenamefont {Wallace},\ and\ \citenamefont {Wilson}}]{Brezin:1972fc}%
  \BibitemOpen
  \bibfield  {author} {\bibinfo {author} {\bibfnamefont {E.}~\bibnamefont
  {Brezin}}, \bibinfo {author} {\bibfnamefont {D.~J.}\ \bibnamefont {Wallace}},
  \ and\ \bibinfo {author} {\bibfnamefont {Kenneth~G.}\ \bibnamefont
  {Wilson}},\ }\bibfield  {title} {\enquote {\bibinfo {title} {{FEYNMAN GRAPH
  EXPANSION FOR THE EQUATION OF STATE NEAR THE CRITICAL POINT (ISING-LIKE
  CASE)}},}\ }\href {\doibase 10.1103/PhysRevLett.29.591} {\bibfield  {journal}
  {\bibinfo  {journal} {Phys. Rev. Lett.}\ }\textbf {\bibinfo {volume} {29}},\
  \bibinfo {pages} {591--594} (\bibinfo {year} {1972})}\BibitemShut {NoStop}%
\bibitem [{\citenamefont {Brezin}\ \emph {et~al.}(1973)\citenamefont {Brezin},
  \citenamefont {Wallace},\ and\ \citenamefont {Wilson}}]{Brezin:1972fb}%
  \BibitemOpen
  \bibfield  {author} {\bibinfo {author} {\bibfnamefont {E.}~\bibnamefont
  {Brezin}}, \bibinfo {author} {\bibfnamefont {D.~J.}\ \bibnamefont {Wallace}},
  \ and\ \bibinfo {author} {\bibfnamefont {Kenneth}\ \bibnamefont {Wilson}},\
  }\bibfield  {title} {\enquote {\bibinfo {title} {{FEYNMAN-GRAPH EXPANSION FOR
  THE EQUATION OF STATE NEAR THE CRITICAL POINT}},}\ }\href {\doibase
  10.1103/PhysRevB.7.232} {\bibfield  {journal} {\bibinfo  {journal} {Phys.
  Rev. B}\ }\textbf {\bibinfo {volume} {7}},\ \bibinfo {pages} {232--239}
  (\bibinfo {year} {1973})}\BibitemShut {NoStop}%
\bibitem [{\citenamefont {Brezin}\ \emph {et~al.}(1974)\citenamefont {Brezin},
  \citenamefont {{Le Guillou}},\ and\ \citenamefont
  {Zinn-Justin}}]{BREZIN1974285}%
  \BibitemOpen
  \bibfield  {author} {\bibinfo {author} {\bibfnamefont {E.}~\bibnamefont
  {Brezin}}, \bibinfo {author} {\bibfnamefont {J-C.}\ \bibnamefont {{Le
  Guillou}}}, \ and\ \bibinfo {author} {\bibfnamefont {J.}~\bibnamefont
  {Zinn-Justin}},\ }\bibfield  {title} {\enquote {\bibinfo {title} {Universal
  ratios of critical amplitudes near four dimensions},}\ }\href {\doibase
  https://doi.org/10.1016/0375-9601(74)90168-6} {\bibfield  {journal} {\bibinfo
   {journal} {Physics Letters A}\ }\textbf {\bibinfo {volume} {47}},\ \bibinfo
  {pages} {285--287} (\bibinfo {year} {1974})}\BibitemShut {NoStop}%
\bibitem [{\citenamefont {Bervillier}(1976)}]{Bervillier:1976zz}%
  \BibitemOpen
  \bibfield  {author} {\bibinfo {author} {\bibfnamefont {C.}~\bibnamefont
  {Bervillier}},\ }\bibfield  {title} {\enquote {\bibinfo {title} {{Universal
  relations among critical amplitude. Calculations up to order epsilon2 for
  systems with continuous symmetry}},}\ }\href {\doibase
  10.1103/PhysRevB.14.4964} {\bibfield  {journal} {\bibinfo  {journal} {Phys.
  Rev. B}\ }\textbf {\bibinfo {volume} {14}},\ \bibinfo {pages} {4964--4975}
  (\bibinfo {year} {1976})}\BibitemShut {NoStop}%
\bibitem [{\citenamefont {Gaiotto}\ \emph {et~al.}(2014)\citenamefont
  {Gaiotto}, \citenamefont {Mazac},\ and\ \citenamefont
  {Paulos}}]{Gaiotto:2013nva}%
  \BibitemOpen
  \bibfield  {author} {\bibinfo {author} {\bibfnamefont {D.}~\bibnamefont
  {Gaiotto}}, \bibinfo {author} {\bibfnamefont {D.}~\bibnamefont {Mazac}}, \
  and\ \bibinfo {author} {\bibfnamefont {M.~F.}\ \bibnamefont {Paulos}},\
  }\bibfield  {title} {\enquote {\bibinfo {title} {{Bootstrapping the 3d Ising
  twist defect}},}\ }\href {\doibase 10.1007/JHEP03(2014)100} {\bibfield
  {journal} {\bibinfo  {journal} {JHEP}\ }\textbf {\bibinfo {volume} {03}},\
  \bibinfo {pages} {100} (\bibinfo {year} {2014})},\ \Eprint
  {http://arxiv.org/abs/1310.5078} {arXiv:1310.5078 [hep-th]} \BibitemShut
  {NoStop}%
\bibitem [{\citenamefont {Gopakumar}\ \emph
  {et~al.}(2017{\natexlab{a}})\citenamefont {Gopakumar}, \citenamefont
  {Kaviraj}, \citenamefont {Sen},\ and\ \citenamefont
  {Sinha}}]{Gopakumar:2016wkt}%
  \BibitemOpen
  \bibfield  {author} {\bibinfo {author} {\bibfnamefont {R.}~\bibnamefont
  {Gopakumar}}, \bibinfo {author} {\bibfnamefont {A.}~\bibnamefont {Kaviraj}},
  \bibinfo {author} {\bibfnamefont {K.}~\bibnamefont {Sen}}, \ and\ \bibinfo
  {author} {\bibfnamefont {A.}~\bibnamefont {Sinha}},\ }\bibfield  {title}
  {\enquote {\bibinfo {title} {{Conformal Bootstrap in Mellin Space}},}\ }\href
  {\doibase 10.1103/PhysRevLett.118.081601} {\bibfield  {journal} {\bibinfo
  {journal} {Phys. Rev. Lett.}\ }\textbf {\bibinfo {volume} {118}},\ \bibinfo
  {pages} {081601} (\bibinfo {year} {2017}{\natexlab{a}})},\ \Eprint
  {http://arxiv.org/abs/1609.00572} {arXiv:1609.00572 [hep-th]} \BibitemShut
  {NoStop}%
\bibitem [{\citenamefont {Gopakumar}\ \emph
  {et~al.}(2017{\natexlab{b}})\citenamefont {Gopakumar}, \citenamefont
  {Kaviraj}, \citenamefont {Sen},\ and\ \citenamefont
  {Sinha}}]{Gopakumar:2016cpb}%
  \BibitemOpen
  \bibfield  {author} {\bibinfo {author} {\bibfnamefont {R.}~\bibnamefont
  {Gopakumar}}, \bibinfo {author} {\bibfnamefont {A.}~\bibnamefont {Kaviraj}},
  \bibinfo {author} {\bibfnamefont {K.}~\bibnamefont {Sen}}, \ and\ \bibinfo
  {author} {\bibfnamefont {A.}~\bibnamefont {Sinha}},\ }\bibfield  {title}
  {\enquote {\bibinfo {title} {{A Mellin space approach to the conformal
  bootstrap}},}\ }\href {\doibase 10.1007/JHEP05(2017)027} {\bibfield
  {journal} {\bibinfo  {journal} {JHEP}\ }\textbf {\bibinfo {volume} {05}},\
  \bibinfo {pages} {027} (\bibinfo {year} {2017}{\natexlab{b}})},\ \Eprint
  {http://arxiv.org/abs/1611.08407} {arXiv:1611.08407 [hep-th]} \BibitemShut
  {NoStop}%
\bibitem [{\citenamefont {{Dey}}\ \emph {et~al.}(2017)\citenamefont {{Dey}},
  \citenamefont {{Kaviraj}},\ and\ \citenamefont {{Sinha}}}]{Dey2017a}%
  \BibitemOpen
  \bibfield  {author} {\bibinfo {author} {\bibfnamefont {Parijat}\ \bibnamefont
  {{Dey}}}, \bibinfo {author} {\bibfnamefont {Apratim}\ \bibnamefont
  {{Kaviraj}}}, \ and\ \bibinfo {author} {\bibfnamefont {Aninda}\ \bibnamefont
  {{Sinha}}},\ }\bibfield  {title} {\enquote {\bibinfo {title} {{Mellin space
  bootstrap for global symmetry}},}\ }\href {\doibase 10.1007/JHEP07(2017)019}
  {\bibfield  {journal} {\bibinfo  {journal} {J. High Energy Phys.}\ }\textbf
  {\bibinfo {volume} {2017}},\ \bibinfo {eid} {19} (\bibinfo {year} {2017})},\
  \Eprint {http://arxiv.org/abs/1612.05032} {arXiv:1612.05032 [hep-th]}
  \BibitemShut {NoStop}%
\bibitem [{\citenamefont {Gopakumar}\ and\ \citenamefont
  {Sinha}(2018)}]{Gopakumar:2018xqi}%
  \BibitemOpen
  \bibfield  {author} {\bibinfo {author} {\bibfnamefont {R.}~\bibnamefont
  {Gopakumar}}\ and\ \bibinfo {author} {\bibfnamefont {A.}~\bibnamefont
  {Sinha}},\ }\bibfield  {title} {\enquote {\bibinfo {title} {{On the
  Polyakov-Mellin bootstrap}},}\ }\href {\doibase 10.1007/JHEP12(2018)040}
  {\bibfield  {journal} {\bibinfo  {journal} {JHEP}\ }\textbf {\bibinfo
  {volume} {12}},\ \bibinfo {pages} {040} (\bibinfo {year} {2018})},\ \Eprint
  {http://arxiv.org/abs/1809.10975} {arXiv:1809.10975 [hep-th]} \BibitemShut
  {NoStop}%
\bibitem [{\citenamefont {Kos}\ \emph {et~al.}(2016)\citenamefont {Kos},
  \citenamefont {Poland}, \citenamefont {Simmons-Duffin},\ and\ \citenamefont
  {Vichi}}]{Kos:2016ysd}%
  \BibitemOpen
  \bibfield  {author} {\bibinfo {author} {\bibfnamefont {Filip}\ \bibnamefont
  {Kos}}, \bibinfo {author} {\bibfnamefont {David}\ \bibnamefont {Poland}},
  \bibinfo {author} {\bibfnamefont {David}\ \bibnamefont {Simmons-Duffin}}, \
  and\ \bibinfo {author} {\bibfnamefont {Alessandro}\ \bibnamefont {Vichi}},\
  }\bibfield  {title} {\enquote {\bibinfo {title} {{Precision Islands in the
  Ising and $O(N)$ Models}},}\ }\href {\doibase 10.1007/JHEP08(2016)036}
  {\bibfield  {journal} {\bibinfo  {journal} {JHEP}\ }\textbf {\bibinfo
  {volume} {08}},\ \bibinfo {pages} {036} (\bibinfo {year} {2016})},\ \Eprint
  {http://arxiv.org/abs/1603.04436} {arXiv:1603.04436 [hep-th]} \BibitemShut
  {NoStop}%
\bibitem [{\citenamefont {Rose}\ \emph {et~al.}(2022)\citenamefont {Rose},
  \citenamefont {Pagani},\ and\ \citenamefont {Dupuis}}]{Rose:2021zdk}%
  \BibitemOpen
  \bibfield  {author} {\bibinfo {author} {\bibfnamefont {F\'elix}\ \bibnamefont
  {Rose}}, \bibinfo {author} {\bibfnamefont {Carlo}\ \bibnamefont {Pagani}}, \
  and\ \bibinfo {author} {\bibfnamefont {Nicolas}\ \bibnamefont {Dupuis}},\
  }\bibfield  {title} {\enquote {\bibinfo {title} {{Operator product expansion
  coefficients from the nonperturbative functional renormalization group}},}\
  }\href {\doibase 10.1103/PhysRevD.105.065020} {\bibfield  {journal} {\bibinfo
   {journal} {Phys. Rev. D}\ }\textbf {\bibinfo {volume} {105}},\ \bibinfo
  {pages} {065020} (\bibinfo {year} {2022})},\ \Eprint
  {http://arxiv.org/abs/2110.13174} {arXiv:2110.13174 [hep-th]} \BibitemShut
  {NoStop}%
\bibitem [{\citenamefont {Berezin}(1980)}]{Berezin:1980xw}%
  \BibitemOpen
  \bibfield  {author} {\bibinfo {author} {\bibfnamefont {F.~A.}\ \bibnamefont
  {Berezin}},\ }\bibfield  {title} {\enquote {\bibinfo {title} {{Feynman Path
  Integrals in a Phase Space}},}\ }\href {\doibase
  10.1070/PU1980v023n11ABEH005062} {\bibfield  {journal} {\bibinfo  {journal}
  {Usp. Fiz. Nauk}\ }\textbf {\bibinfo {volume} {132}},\ \bibinfo {pages}
  {497--548} (\bibinfo {year} {1980})}\BibitemShut {NoStop}%
\bibitem [{\citenamefont {Moshe}\ and\ \citenamefont
  {Zinn-Justin}(2003)}]{Moshe:2003xn}%
  \BibitemOpen
  \bibfield  {author} {\bibinfo {author} {\bibfnamefont {Moshe}\ \bibnamefont
  {Moshe}}\ and\ \bibinfo {author} {\bibfnamefont {Jean}\ \bibnamefont
  {Zinn-Justin}},\ }\bibfield  {title} {\enquote {\bibinfo {title} {{Quantum
  field theory in the large N limit: A Review}},}\ }\href {\doibase
  10.1016/S0370-1573(03)00263-1} {\bibfield  {journal} {\bibinfo  {journal}
  {Phys. Rept.}\ }\textbf {\bibinfo {volume} {385}},\ \bibinfo {pages}
  {69--228} (\bibinfo {year} {2003})},\ \Eprint
  {http://arxiv.org/abs/hep-th/0306133} {arXiv:hep-th/0306133} \BibitemShut
  {NoStop}%
\bibitem [{\citenamefont {Hohenberg}\ and\ \citenamefont
  {Halperin}(1977)}]{Hohenberg:1977ym}%
  \BibitemOpen
  \bibfield  {author} {\bibinfo {author} {\bibfnamefont {P.~C.}\ \bibnamefont
  {Hohenberg}}\ and\ \bibinfo {author} {\bibfnamefont {B.~I.}\ \bibnamefont
  {Halperin}},\ }\bibfield  {title} {\enquote {\bibinfo {title} {{Theory of
  Dynamic Critical Phenomena}},}\ }\href {\doibase 10.1103/RevModPhys.49.435}
  {\bibfield  {journal} {\bibinfo  {journal} {Rev. Mod. Phys.}\ }\textbf
  {\bibinfo {volume} {49}},\ \bibinfo {pages} {435--479} (\bibinfo {year}
  {1977})}\BibitemShut {NoStop}%
\bibitem [{\citenamefont {Berges}(2015)}]{Berges:2015kfa}%
  \BibitemOpen
  \bibfield  {author} {\bibinfo {author} {\bibfnamefont {Jurgen}\ \bibnamefont
  {Berges}},\ }\bibfield  {title} {\enquote {\bibinfo {title} {{Nonequilibrium
  Quantum Fields: From Cold Atoms to Cosmology}},}\ }\href@noop {} {\
  (\bibinfo {year} {2015})},\ \Eprint {http://arxiv.org/abs/1503.02907}
  {arXiv:1503.02907 [hep-ph]} \BibitemShut {NoStop}%
\bibitem [{\citenamefont {Chaichian}\ and\ \citenamefont
  {Demichev}(2001)}]{Chaichian:2001cz}%
  \BibitemOpen
  \bibfield  {author} {\bibinfo {author} {\bibfnamefont {M.}~\bibnamefont
  {Chaichian}}\ and\ \bibinfo {author} {\bibfnamefont {A.}~\bibnamefont
  {Demichev}},\ }\href@noop {} {\emph {\bibinfo {title} {{Path integrals in
  physics. Vol. 1: Stochastic processes and quantum mechanics}}}}\ (\bibinfo
  {publisher} {{Institute of Physics Publishing}},\ \bibinfo {year}
  {2001})\BibitemShut {NoStop}%
\end{thebibliography}%

\end{document}